# PSFHS Challenge Report: Pubic Symphysis and Fetal Head Segmentation from Intrapartum Ultrasound Images

Jieyun Bai [a,b,1,2*], Zihao Zhou [a,1], Zhanhong Ou [a,1,2], Gregor Koehler [c], Raphael Stock [c], Klaus Maier-Hein [c], Marawan Elbatel [d], Robert Martí [e], Xiaomeng Li [d], Yaoyang Qiu [f], Panjie Gou [f], Gongping Chen [g], Lei Zhao [h], Jianxun Zhang [g], Yu Dai [g], Fangyijie Wang [i], Guénolé Silvestre [i], Kathleen Curran [j], Hongkun Sun [k], Jing Xu [k], Pengzhou Cai [l], Lu Jiang [l], Libin Lan [l], Dong Ni [m,2], Mei Zhong [n,2], Gaowen Chen [o,2], Víctor M. Campello [p,2], Yaosheng Lu [a,2*] and Karim Lekadir [p,q,2]

[a] Guangdong Provincial Key Laboratory of Traditional Chinese Medicine Informatization, Jinan University, Guangzhou, China

[b] Auckland Bioengineering Institute, The University of Auckland, Auckland, New Zealand

[c] Division of Medical Image Computing, German Cancer Research Center (DKFZ), Heidelberg, Germany

[d] Department of Electronic and Computer Engineering, The Hong Kong University of Science and Technology, Hongkong, China

[e] Computer Vision and Robotics Group, University of Girona, Girona, Spain

[f] Canon Medical Systems (China) Co., LTD, Beijing, China

[g] College of Artificial Intelligence, Nankai University, Tianjin, China

[h] College of Computer Science and Electronic Engineering, Hunan University, Changsha, China

[i] School of Medicine, University College Dublin, Dublin, Ireland

[j] School of Computer Science, University College Dublin, Dublin, Ireland

[k] School of Statistics & Mathematics, Zhejiang Gongshang University, Hangzhou, China

[l] School of Computer Science & Engineering, Chongqing University of Technology, Chongqing, China

[m] National-Regional Key Technology Engineering Laboratory for Medical Ultrasound & Guangdong Provincial Key Laboratory of Biomedical Measurements and Ultrasound Imaging & School of Biomedical Engineering, Health Science Center, Shenzhen University, Shenzhen, China

[n] NanFang Hospital of Southern Medical University, Guangzhou, China

[o] Zhujiang Hospital of Southern Medical University, Guangzhou, China

[p] Departament de Matemàtiques i Informàtica, Universitat de Barcelona, Barcelona, Spain

[q] Institució Catalana de Recerca i Estudis Avançats (ICREA), Barcelona, Spain

[1] These authors contributed equally to the work.
[2] These authors co-organized the PSFHS challenge. All others contributed results of their algorithms presented in the paper.
*Corresponding author at Auckland Bioengineering Institute, The University of Auckland, Private Bag 92019, Auckland 1142, New Zealand
E-mail address: jbai996@aucklanduni.ac.nz (Jieyun Bai) and tluys@jnu.edu.cn (Yaosheng Lu)

# Abstract

Segmentation of the fetal and maternal structures, particularly intrapartum ultrasound imaging as advocated by the International Society of Ultrasound in Obstetrics and Gynecology (ISUOG) for monitoring labor progression, is a crucial first step for quantitative diagnosis and clinical decision-making. This requires specialized analysis by obstetrics professionals, in a task that i) is highly time- and cost-consuming and ii) often yields inconsistent results. The utility of automatic segmentation algorithms for biometry has been proven, though existing results remain suboptimal. To push forward advancements in this area, the Grand Challenge on Pubic Symphysis-Fetal Head Segmentation (PSFHS) was held alongside the 26th International Conference on Medical Image Computing and Computer Assisted Intervention (MICCAI 2023). This challenge aimed to enhance the development of automatic segmentation algorithms at an international scale, providing the largest dataset to date with 5,101 intrapartum ultrasound images collected from two ultrasound machines across three hospitals from two institutions. The scientific community's enthusiastic participation led to the selection of the top 8 out of 179 entries from 193 registrants in the initial phase to proceed to the competition's second stage. These algorithms have elevated the state-of-the-art in automatic PSFHS from intrapartum ultrasound images. A thorough analysis of the results pinpointed ongoing challenges in the field and outlined recommendations for future work. The top solutions and the complete dataset remain publicly available, fostering further advancements in automatic segmentation and biometry for intrapartum ultrasound imaging.

**Keywords:** Intrapartum ultrasound; fetal biometry; deep learning; challenge; angle of progress; image segmentation

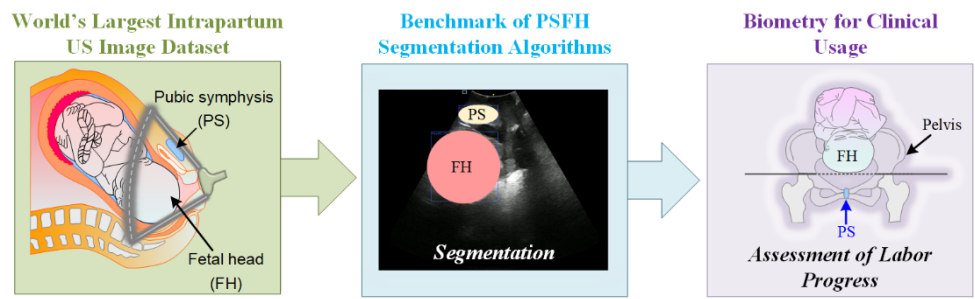

## Highlights

- Established a benchmark for segmentation algorithms of fetal and maternal structures.
- Utilized the first and largest publicly available intrapartum ultrasound image dataset.
- The 'Segment Anything Model' demonstrated superior segmentation performance.
- Released the challenge algorithms and the test dataset for ongoing research and exploration.

## 1. Introduction

Intrapartum ultrasound imaging of fetal and maternal structures (specifically, pubic symphysis and fetal head, PSFH) provides clinicians and researchers with critical visual insights into labor progression, which involves tracking the descent of the fetus relative to the maternal pelvis (Sherer, 2007). Ultrasound imaging technique is invaluable not only during the first and second stages of labor but continues to be useful up until birth. It plays a significant role in the management of abnormal labor (Gimovsky, 2021), and in advising the optimal timing for labor induction (Badr et al., 2024). Intrapartum ultrasound is instrumental in predicting labor duration (Carvalho Neto et al., 2021), delivery method, and the likelihood of complicated operative interventions (Hadad et al., 2021), with the angle of progression (AoP)—both at rest and its variations—serving as key predictors (Angeli et al., 2020) (**Figure 1**).

Automated segmentation and quantification of the complex, dynamically changing PSFH during labor can significantly enhance diagnostic accuracy, given that manual segmentation is both time-intensive and prone to human error and inter-rater variability (Pietsch et al., 2021). It is clinically pertinent to analyze the morphometry of the evolving PSFH, where measures like the AoP can be objectively assessed against population-based benchmarks of normal progression (Angeli et al., 2020). The AoP is defined as the angle between the longitudinal axis of the pubic bone and a line extending from the lower edge of the pubic symphysis to tangentially touch the deepest bony part of the fetal head (Kalache et al., 2009). Research indicates that an AoP of 120° or more is strongly associated with a high likelihood of spontaneous vaginal delivery (Dall'Asta et al., 2019). Currently, AoP measurements are predominantly performed manually or semi-automatically (Conversano et al., 2017; Haberman et al., 2021), with full automation of PSFH segmentation for AoP biometry remaining a largely unexplored domain.

Technical challenges abound for automatic segmentation of PSFH during labor, which sees significant anatomical changes due to uterine contractions that affect image clarity and alter spatial relationships between the fetal head and pubic symphysis (Sharf et al., 2007). The smaller size of the pubic symphysis compared to the fetal head (Pavličev et al., 2020) adds complexity to precise segmentation, crucial for accurate AoP biometry. The measurement of AoP involves three critical landmarks—two associated with the pubic symphysis and one with the fetal head—demonstrating significant dependency in this segmentation process **(Figure 1A-B)**. Ultrasound imaging itself presents challenges; it is patient-specific, operator-dependent, and machine-specific (Zhao et al., 2023). The intrinsic properties of ultrasound, such as signal dropouts, artefacts, missing boundaries, attenuation, shadows, and speckle, add to the modality's complexity (Li et al., 2022; Wright et al., 2023; Zamzmi et al., 2022; Zimmer et al., 2023). Depending on the transducer's orientation, images may appear distorted or incomplete, presenting additional challenges for automatic methods (Mischi et al., 2020; Torres et al., 2022).

Recent advancements underscore the potential for automatic methods to transform this field (Bai et al., 2024; Baumgartner et al., 2017; Carneiro et al., 2008; Chen et al., 2017; Fiorentino et al., 2023; Jang et al., 2018; Lin et al., 2019; Pu et al., 2021; Wang et al., 2022b; Wu et al., 2017; Yang et al., 2019; Yu et al., 2018). Since the International Society of Ultrasound in Obstetrics and Gynecology (ISUOG) introduced the Practice Guidelines for Intrapartum Ultrasound in 2018 (Ghi et al., 2018; Vogel et al., 2024), various deep learning strategies, especially those based on UNet architectures (LeCun et al., 2015), have been developed for the segmentation of pubic symphysis and fetal head (Bai et al., 2022; Chen et al., 2024d; Lu et al., 2022a; Ou et al., 2024). Innovations include dual attention decoders and dual decoder strategies, which improve feature extraction and boundary delineation in intrapartum ultrasound images (Chen et al., 2024b; Chen et al., 2024c). Further enhancements incorporate shape-constrained loss functions and directional information to optimize model performance. However, the evaluation and broad application of these methodologies are limited by the availability of diverse and large datasets (Chen et al., 2024a; Lu et al., 2022b). Therefore, the development and dissemination of annotated datasets from multiple devices and centers are crucial for advancing automated tools in intrapartum ultrasound imaging, facilitating robust computer-aided diagnostic systems to support labor and delivery management.

This paper details the organization of the Pubic Symphysis-Fetal Head Segmentation (PSFHS) Challenge, describes the submitted segmentation frameworks, and provides a thorough evaluation of the results using the Biomedical Image Analysis ChallengeS (BIAS) method (Maier-Hein et al., 2020). The PSFHS Challenge aimed to develop reliable, valid, and reproducible methods for analyzing intrapartum ultrasound images of PSFH during labor, using a dataset from two institutions and two vendors to drive the development of automatic segmentation methods for AoP biometry. Our evaluation compares algorithms on two concealed testing datasets, examining performance variations across different hospitals, ultrasound machines, and relative positions (i.e., AoP≥120° and AoP<120°) (**Figure 1C**). We also explored the impacts of architectural design, data preprocessing, post-processing, loss functions, and optimizers on segmentation performance. Our comprehensive benchmarking study offers insights into necessary design choices and practical considerations. Through this global benchmarking effort, we aim to optimize the framework for PSFH segmentation for AoP biometry, providing critical insights into effective strategies and techniques. These algorithms developed through the PSFHS Challenge will enhance our understanding of the underlying causes of labor arrest and guide the development of intrapartum guidelines and clinical risk stratification tools for early interventions, treatments, and care management decisions (**Figure 1D**). These approaches could potentially extend to other areas of fetal segmentation and obstetric ultrasound imaging, having a significant impact on the broader imaging community.

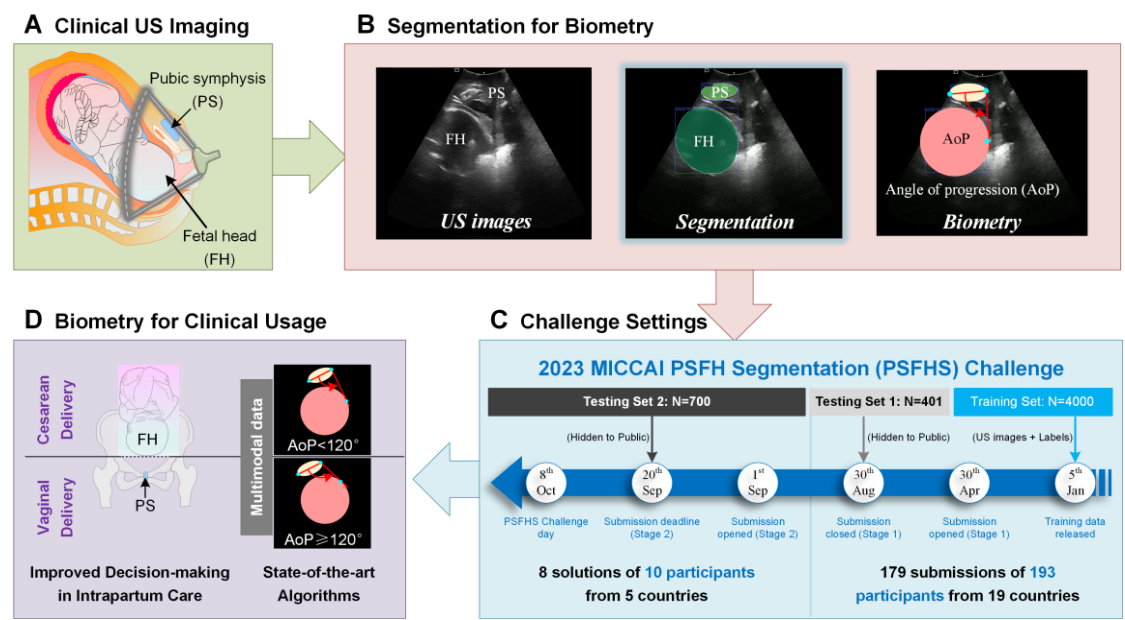

**Figure 1. Overall Workflow of Clinical Image Utilization and the Pubic Symphysis-Fetal Head (PSFH) Segmentation Challenge. A)** Clinical images are acquired via a transperineal ultrasound (US) approach using mid-sagittal scans from pregnant women during labor. **B)** Expert manual segmentation of US images is performed to delineate the Pubic Symphysis (PS) and Fetal

Head (FH). This facilitates the derivation of the biometric parameter—angle of progression (AoP), through the identification of three critical landmarks within the PSFH segmentation. **C)** In the PSFHS Challenge, a total of 5,101 2D US images were categorized, with 4,000 designated for training, 401 for initial testing (Testing Set 1), and 700 for final evaluation (Testing Set 2). The training dataset was made accessible to all registered participants, while Testing Set 1 was used to refine and select the top 10 methodologies. Testing Set 2 was utilized for the conclusive assessment of these methodologies. The eight top-performing algorithms, along with their source codes, were subsequently ranked based on segmentation and measurement metrics. **D)** Accurate measurement of the AoP, based on the anatomical structures of the PSFH, provides crucial information with other multimodal data for assessing labor progression and predicting the mode of delivery.

## 2. Material and Methods

2.1 Challenge setup

The Pubic Symphysis-Fetal Head Segmentation (PSFHS) Challenge was hosted alongside the 26th International Conference on Medical Image Computing and Computer Assisted Intervention (MICCAI 2023) on the Grand Challenge platform (https://ps-fh-aop-2023.grand-challenge.org/) to facilitate the comparison of different algorithms in the field of intrapartum ultrasound image analysis. It is a repeated event with annual MICCAI submission deadline. The challenge targeted the segmentation of the pubic symphysis (PS) and fetal head (FH) regions using deep learning techniques.

The challenge was organized by a subgroup specializing in "Deep Learning in Intrapartum Ultrasound Image Analysis," part of a broader initiative focused on the industrialization of key technologies for labor monitoring. The organizing team included multidisciplinary experts such as early-stage researchers, PhDs, assistant, associate, and full professors, along with practicing obstetricians and sonographers from three medical centers of two universities.

In the challenge, members of the organizer' institutes can participate but not eligible for awards and not listed in leaderboard. These participants were tasked to develop and refine machine learning models for segmenting PSFH from the intrapartum ultrasound images. They were allowed to use pre-trained models to provide fully automated segmentation methods. The challenge unfolded in two stages:

- **Training and Initial Submission:** Commencing on January 1, 2023, participants accessed the training dataset via Zenodo (available at https://zenodo.org/records/7851339#.ZEH6eHZBztU). Competitors were required to develop their algorithms and encapsulate them within Docker containers. These containers were subsequently submitted to the grand-challenge platform, where their performance was assessed using a first set of test images, which remained undisclosed to the participants. This evaluation phase concluded on August 30, 2023.
- **Final Evaluation:** The ten highest-performing teams from the initial phase advanced to the second stage. Modifications were necessitated due to overlapping methodologies, specifically, two teams from the same institution employing identical strategies were combined, and another was disqualified for not providing sufficient details about their algorithm. This adjustment left eight teams in the competition. These teams submitted their revised Docker containers for final evaluation against a second set of concealed test images by September 20, 2023. The top seven teams were honored during an awards ceremony on October 8, 2023, the recording of which is available at https://www.youtube.com/watch?v=NvP-bT8I1fE&t=1s. All the teams were invited to contribute to draft and submit a manuscript describing the methods and results. The first and last author of these teams were listed as authors of this challenge report. The participating teams also can publish their own results separately after coordination to avoid significant overlap with the challenge report.

All algorithms developed as part of this challenge have been made publicly available on GitHub (https://github.com/maskoffs/PS-FH-MICCAI23). For a detailed overview of the challenge setup and objectives, please refer to the final challenge proposal documented on Zenodo (https://zenodo.org/records/7861699).

2.2 Dataset

**Dataset Overview.** The PSFHS2023 dataset consists of 5,101 2D intrapartum ultrasound images retrospectively collected from 1,175 pregnant women with a variety of age (from 18- to 46-year-old) to evaluate the fetal head station at the onset of the second stage of labor using transperineal ultrasound. These images were sourced from three medical institutions affiliated with two universities: Nanfang Hospital and Zhujiang Hospital (both at Southern Medical University), and The First Affiliated Hospital of Jinan University (**Figure 2A**). The inclusion criteria were: singleton pregnancy, at least 37 + 0 weeks of pregnancy and longitudinal cephalic fetus presentation. The exclusion criteria were: preterm delivery, other than longitudinal cephalic fetus presentation; multiple pregnancy, uterine abnormalities, condition after uterus surgery, pathological intrapartum cardiotocography and the refusal of the patient to participate in the study.

**Data Division in the Challenge.** The dataset is organized into three subsets: the Training Set, containing 4,000 images from 305 patients; Testing Set 1, which includes 401 images from 325 patients; and Testing Set 2, consisting of 700 images from 545 patients (**Table 1**).

**Ethical Compliance.** The data collected for this challenge received approval from the institutional review boards of Zhujiang Hospital of Southern Medical University (No. 2023-SYJS-023), Nanfang Hospital of Southern Medical University (No. NFCE-2019–024) and the First Affiliated hospital of Jinan University (No. JNUKY-2022-019). Informed consent was waived because of the retrospective nature of the study and the analysis used anonymous medical image data. Informed consent was waived because of the retrospective nature of the study and the analysis used anonymous medical image data. The data is available under the CC BY license, with additional details accessible on Zenodo: the Training set at https://zenodo.org/records/7851339, and Testing Sets 1 and 2 at https://zenodo.org/records/10969427.

**Imaging Equipment and Protocols.** All the images from this study were acquired by trained clinicians using Esaote MyLab (Esaote SpA in Italy) or ObEye (Guangzhou Lianyin Medical Technology Co., Ltd. in China) and following the protocols defined by the International Society of Ultrasound in Obstetrics & Gynecology guidelines. The transducer was prepped by covering it with a surgical latex glove filled with coupling gel, then the prepped transducer, after applying gel, was placed between labia below the pubic symphysis to obtain a sagittal plane, small adjustments in the form of lateral movements of the probe were made until an image obtained showed clear maternal pelvic (pubic symphysis) and fetal (fetal skull) landmarks that did not show any shadows from the pubic rami. Most of the images were acquired with a 3.53±0.0525MHz convex probe. The spatial resolution of the ultrasound system is specified by the manufacturer to less than 2 mm. The overall geometric inaccuracy of a very similar setup due to inherent technical limitations was measured to be <2.0 mm laterally, <2.0 mm vertically, <2.0 mm longitudinally, and <8.0 mm radially ('vector length' or Euclidean '3D-distance'; the square root of the sum of squares of the three axes) consisting of random errors (per single measurement point) and systematic errors (effectively, per fraction). The temporal resolution of the device is specified to about 27 Hz. These images were in BMP format, anonymised, and automatically cropped (to remove the header) to a size of 256×256 pixels before distribution. Spatial resolution (in millimeters) varied among the images. Image distribution per device is as follows: the Training Set contains 3,743 images from Esaote MyLab and 257 from ObEye; Testing Set 1 includes 100 images from Esaote MyLab and 301 from ObEye; and Testing Set 2 consists of 213 images from Esaote MyLab and 487 from ObEye (**Table 1**).

**Image Annotation and Quality Assessment.** The team responsible for annotations included two proficient physicians and 18 students specializing in biomedical studies. Before commencing their tasks, annotators received comprehensive training that involved familiarizing them with the structures of PSFH and the key aspects of ultrasound images. This training was facilitated through a combination of online sessions and in-person guidance by the physicians. Each annotator was assigned 15 test images, which were subsequently assessed by the physicians. If the annotations were deemed inadequate, the images were returned to the respective student for refinement. Annotators were instructed to utilize the pencil tool in Pair (https://www.aipair.com.cn/) for precise pixel-wise segmentation. In instances where the contours appeared fragmented or discontinuous, annotators were instructed to ensure that the contours maintained a complete elliptical shape. This instruction was essential considering the ultimate clinical application's requirement to calculate AoP based on the segmented PSFH contours. The final segmentation ground truth was represented by a three-color image, where red pixels denoted PS, green pixels represented FH, and black pixels indicated the background. During the official annotation phase, each image was annotated by two annotators. Any overlapping pixels annotated by both annotators were further reviewed and adjusted by a highly experienced physician with a decade of expertise. Here, the non-overlapping pixels account for >5% of the total labelled pixels annotated by two annotators, and we believe that physicians need to make modifications. Note: Dropped artifacts in ultrasound images can significantly impact the quality and accuracy of the images, leading to potential errors in annotation. To evaluate the quality of segmentation, both intra-annotator and inter-annotator variabilities were assessed using 40 images, each annotated twice by three different annotators (including one experienced clinician and two trained raters) on separate occasions. These variabilities were quantified using the PSFH Dice score calculated with respect to ground truths of our dataset. (**Table 2**).

**Data Covers All Scenarios of Fetal Head Stations.** Fetal head station can be measured objectively by AoP to assess current status and as a baseline for longitudinal measurements. Head station should be assessed transperineally, not transabdominally. AoP (in degrees) ranging from 84° to 170° is equivalent to head station expressed in centimeters, from –3 cm to +5 cm (direct conversion is possible), and has the potential to link ultrasound data to traditional assessment by palpation. (Tutschek et al., 2013). It can also help to predict whether operative vaginal delivery is likely to be successful. An AoP≥120° significantly correlates with a high probability of spontaneous vaginal delivery (Dall'Asta et al., 2019). The AoP distribution is structured as follows: AoPs range from 50° to 180° and the ratio of AoP≥120°/AoP<120° is 269/3731 in the Training Set (**Figure 2B**); AoPs range from 60° to 160° and the ratio of AoP≥120°/AoP<120° is 116/285 in the Testing Set 1 (**Figure 2C**); and AoPs range from 60° to 180° and the ratio of AoP≥120°/AoP<120° is 122/578 in the Testing Set 2 (**Figure 2D**) (**Table 1**). This detailed dataset structure allows for robust training and evaluation of automatic segmentation algorithms in real-world clinical applications.

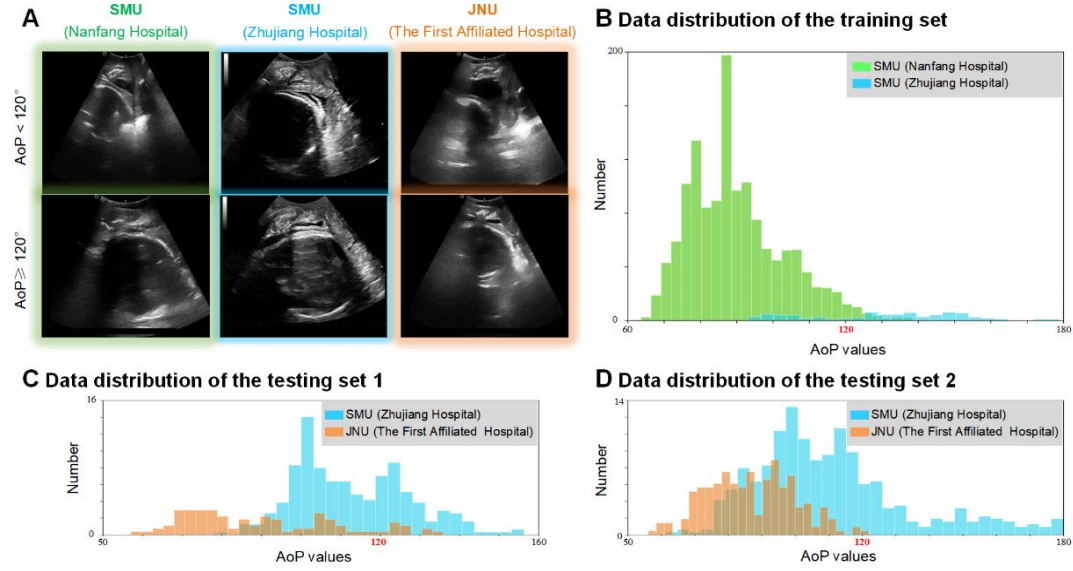

**Figure 2.** Data Sources and Distribution of the US Image Dataset Used for This Study. **A)** Representative cases (i.e., AoP<120° and AoP≥120°) from each hospital in the US image dataset are illustrated. Images from the Nanfang Hospital of Southern Medical University (SMU) and the First Affiliated Hospital of Jinan University (JNU) were captured using the ObEye system. In contrast, those from the Zhujiang Hospital of SMU were captured with the Esaote MyLab system. **B)** The training dataset includes images sourced from SMU (Nanfang Hospital and Zhujiang Hospital), with AoP values ranging from 60° to 180°. **C)** Testing Set 1 comprises images from SMU (Zhujiang Hospital) and JNU (First Affiliated Hospital), featuring AoP values from 50° to 160°. **D)** Testing Set 2 contains images from SMU (Zhujiang Hospital) and JNU (First Affiliated Hospital), with AoP values extending from 50° to 180°, offering a broad spectrum for comprehensive evaluation.

**Table 1.** Training and Testing dataset properties from all image centres.

| Data sets | Institution | Hospital | Scanner | Images | Subjects | AoP |
|---|---|---|---|---|---|---|
| Training | SMU | Nanfang Hospital | ObEye | 3743 | 51 | AoP<120° (n=3652) <br> AoP≥120° (n=91) |
|  | SMU | Zhujiang Hospital | Esaote My Lab | 257 | 254 | AoP<120° (n=79) <br> AoP≥120° (n=178) |
| Testing Set 1 | SMU | Zhujiang Hospital | Esaote My Lab | 301 | 299 | AoP<120° (n=196) <br> AoP≥120° (n=105) |
|  | JNU | First Affiliated Hospital | ObEye | 100 | 26 | AoP<120° (n=89) <br> AoP≥120° (n=11) |
| Testing Set 2 | SMU | Zhujiang Hospital | Esaote My Lab | 487 | 487 | AoP<120° (n=366) <br> AoP≥120° (n=121) |
|  | JNU | First Affiliated Hospital | ObEye | 213 | 58 | AoP<120° (n=212) <br> AoP≥120° (n=1) |

**Table 2.** PSFH Dice scores calculated with respect to ground truths of our dataset.

|  | Clinician | Trained rater 1 | Trained rater 2 |
|---|---|---|---|
| 1st | 89.04±0.07 | 87.36±0.05 | 85.20±0.17 |
| 2nd | 90.02±0.12 | 88.86±0.05 | 88.58±0.12 |

2.3 Participating teams

A total of 193 participants (from 19 countries) spanning both industry and academia registered for this challenge, initially submitting 179 solutions on Testing Set 1. The top ten participants were invited to compete in an on-site completion on Testing Set 2. Subsequently, we received detailed descriptions of algorithms from 8 teams:

- **angle_avengers:** Köhler et al. implemented a solution based on nnU-Net(Isensee et al., 2021), a versatile framework known for its adaptability across various datasets and consistent top results in semantic segmentation. They enhanced nnU-Net by incorporating residual connections within the encoder and introducing an advanced data augmentation strategy. This strategy not only improves existing techniques but also includes innovative augmentation methods. Their experimental setup processed full-resolution 2D images through a standard network configuration with a batch size of 49 and a patch size of 256×256 pixels, using z-score normalization. They trained over 1000 epochs, building a strong foundation for further enhancements. The culmination of their methodology was an ensemble of models developed through 5-fold cross-validation, designed to deliver refined predictions, adhering strictly to the default nnU-Net protocol and necessary image format conversions.

- **Aloha:** Elbatel et al. adapted the Segment Anything Model (SAM)(Kirillov et al., 2023) specifically for enhancing ultrasound imaging by increasing resolution to 512×512 pixels(Huang et al., 2024; Ma et al., 2024; Mazurowski et al., 2023). They enriched the training dataset using a variety of data augmentation techniques from the MONAI library, including horizontal flipping, Gaussian noise, blurring, random zooming, and affine transformations. The team employed a low-rank (LoRA) fine-tuning strategy using a pre-trained Vision Transformer (ViT-h) model from the SA-1B natural imaging dataset, with encoder parameters fixed during training. They incorporated a warm-up phase and used the AdamW optimizer to adjust a weighted combination of cross-entropy and dice losses, favoring dice loss with a weight of 0.8 over 400 epochs per fold. For inference, they utilized the encoder from the frozen ViT-h model and applied horizontal flipping as test-time augmentation to double the prediction count per model. The methodology involved alternating between five sets of trained LoRA parameters to produce ten predictions per image, which were then combined using a soft ensemble method to enhance robustness. The final post-processing step focused on isolating the largest connected component from each detected object.

- **QiuYaoyang:** Qiu and Guo utilized UperNet with a ResNet101 backbone for advanced image segmentation tasks (Ruiping et al., 2024). UperNet's architecture, which includes a Feature Pyramid Network (FPN) and a Pyramid Pooling Module (PPM), is designed for multi-scale feature integration and global contextual information synthesis. This setup significantly enhances the model's ability to handle complex scenes and objects. Additionally, their network features a decoder that refines and upsamples features to improve segmentation precision. In their preprocessing strategy, Qiu and Guo normalized the data and utilized a range of augmentation techniques, such as random rotation, resizing, horizontal flipping, Gaussian noise addition, and blurring. They introduced Weighted Dice loss, focusing on image boundary areas to improve segmentation accuracy. To augment the model's generalization capabilities, they trained on a large dataset of 4000 images, developing a robust and effective segmentation model capable of handling various imaging conditions.

- **NKCGP:** Chen et al. enhanced the U-net architecture specifically for complex ultrasound image analysis (Chen et al., 2023) by increasing its depth to 15 layers, which includes seven down-sampling and seven up-sampling stages. Within this architecture, each convolutional module consists of two 3x3 convolution layers, followed by batch normalization and LeakyReLU activation. The number of filters in these layers scales symmetrically from 64 to 1024 and then reduces back to 64, ensuring a balanced distribution of parameters throughout the network. A significant enhancement in their design is the integration of a squeeze-and-excitation (SE) block between the encoder and decoder, which recalibrates the feature maps to emphasize more relevant image areas, thus boosting the model's contextual sensitivity. Additionally, they implemented deep supervised constraints during the decoding phase to increase the accuracy of the output prediction masks. For their training strategy, Chen et al. utilized binary cross entropy as the loss function and the Adam optimizer with an initial learning rate of 1e-3. They conducted training over 50 epochs with a batch size of 12, incorporating cross-validation with 20% of the data reserved for validation. This comprehensive training regimen was designed to enhance model performance and ensure effective generalization across various ultrasound imaging conditions.

- **UCD_Med:** Wang et al. adopted a novel approach by coupling a U-Net architecture with a pre-trained Mix Transformer encoder(Wang et al., 2022a), specifically the MiT-B0 model derived from the ImageNet dataset, to advance image segmentation capabilities. Their preprocessing regimen included pixel normalization and an array of data augmentation techniques aimed at enhancing model robustness and generalizability. These techniques encompassed random rotation within a range of -25° to 25°, vertical flipping with a probability of 0.3, horizontal flipping with a probability of 0.5, and coarse dropout, also set at a probability of 0.5. During the training phase, the focus was on minimizing a weighted sum of cross-entropy and Dice losses, which effectively balances the trade-off between class imbalance handling and boundary precision. The training was conducted over 20 epochs, with each batch comprising 10 images, and utilized a steady learning rate of 1e-4. This strategy was carefully designed to optimize the integration of the transformer's capabilities with the spatial decoding strengths of U-Net, ensuring detailed and accurate segmentation outcomes.

- **Kunkunkk:** Sun et al. refined the TransUnet architecture (Lin et al., 2022) by incorporating an Atrous Spatial Pyramid Pooling (ASPP) structure and utilizing a ResNet50 backbone. This advanced architecture starts with a Transformer-based encoder that captures rich global contextual information, paired with a CNN-based decoder, typically a U-Net, which aids in the precise localization of features. The integration of the ASPP module between the encoder and decoder is crucial, as it allows for the capture of multi-scale spatial information through atrous convolutions. This adaptation enhances the model's ability to handle a diverse range of object sizes and shapes, thereby improving versatility across various imaging contexts. For data augmentation, they implemented techniques such as random rotation and flipping. The training process focused on minimizing both cross-entropy and Dice losses over 150 epochs, with each batch containing 8 images. A steady learning rate of 0.01 was maintained throughout the training to ensure optimal convergence and stable progression of learning metrics. This comprehensive methodology not only boosts segmentation accuracy but also strengthens the model's adaptability to input data variations.

- **CQUT-Smart:** Cai et al. introduced the BRAU-Net architecture (Zhu et al., 2023), a novel vision transformer model that incorporates a four-stage pyramid structure. At the core of this architecture is the Bi-Level Routing Attention (BRA), which substantially enhances the model to analyze complex visual data. The architecture begins with overlapped patch embedding and employs patch merging in later stages to progressively reduce the spatial resolution while increasing the channel count. Each stage of the BRAU-Net is composed of BiFormer blocks, which are designed for efficient feature processing. These blocks feature a 3×3 depthwise convolution to encode relative positions, a BRA module for targeted location-based attention, and a two-layer MLP for in-depth relationship modeling and feature embedding. The design of the BRA module is crucial for refining the attention mechanism, enabling dynamic adjustment to the spatial hierarchies within the image data. For optimization, the Adam optimizer was chosen for its effectiveness with sparse gradients and its adaptive learning rate capabilities, which are well-suited for training sophisticated models like the BRAU-Net. To enhance the diversity of the training dataset and improve model generalizability, they implemented data augmentation techniques including flips and rotations. A distinctive feature of their approach is the tailored attention mechanism within the BRAU-Net, which varies the number of attention heads at each stage (2, 4, 8, 16) and sets specific 'topk' values, optimizing the model's focus and efficiency at different levels of abstraction.

- **aspirerabbit**: Ou et al. utilized the U-net architecture (Ou et al., 2024), known for its distinctive U-shaped configuration, comprising a contraction path (encoder) and an expansion path (decoder). The encoder reduces dimensionality through successive downsampling, capturing essential contextual information, while the decoder employs upsampling to enhance localization accuracy. Integral to this architecture are the skip connections that bridge corresponding layers across the encoder and decoder, preserving high-resolution features. The model was trained using the Dice loss function to refine segmentation accuracy. For validation, the dataset was split into five folds (F1-F5), with F5 reserved as the test set and F1-F4 used for cross-validation during training. Model evaluation focused on achieving the highest Dice score in the validation set before subsequent testing on the test set. The implementation leveraged the PyTorch framework and utilized the Adam optimizer, initialized with a learning rate of 0.0001 and a momentum of 0.9. A cosine annealing scheduler adjusted the learning rate, with a minimum limit set at 0.000001. The batch size was maintained at 4 throughout the 200 epochs of training. Images were resized to 256×256 pixels and subjected to random rotations from -30º to 30º, scaling, and normalization, preparing the data for effective training.

Each algorithm is summarized in **Table 3**. All teams submitted a deep learning-based method, most of which were variants based on the U-Net architecture (N = 5). The top two teams used complex models, where 'angle_avengers' team used nnUNet and 'Aloha' team used Segment Anything Model (SAM). Six of the eight (excluding 'angle_avengers' and 'QiuYaoyang') used Dice loss, and top four teams used an ensemble learning method. Every method used a patch size of 256×256 except for one team ('Aloha') who up-sampled the images to 512×512 for ensuring compatibility with SAM. Most submissions (6 of 8) used the AdamW/Adam optimizer. Half of the submissions used pretrained weights to initialize network parameters and cross-validation. A variety of different data augmentation strategies were used, and only one team did not employ data augmentation at all. Four teams employed pre-trained network backbones trained on publicly available datasets, and used normalization to preprocess images (**Table 4**).

2.4 Evaluation

We used Dice similarity coefficient (DSC), Average Surface Distance (ASD) and Hausdorff distance (HD) to evaluate the segmentation results of PS, FH and PSFH, respectively (Maier-Hein et al., 2024).

**DSC** measures the amount of overlap between the manual segmentation (MS) label and the predicted segmentation (PS) result, and is defined as

$$DSC = \frac{2|MS \cap PS|}{|MS| + |PS|}$$

**ASD** provides a comprehensive measure of the surface distance between two objects, capturing both local and global discrepancies in shape and location. It is defined as

$$\text{ASD} = \frac{D(S_1, S_2) + D(S_2, S_1)}{2}$$

where $S_1$ and $S_2$ represent the surfaces of two objects being compared, and $D(S_1, S_2)$ denotes that the average of these shortest distances for all points on $S_1$, which gives the distance from $S_1$ to $S_2$.

**HD** is a spatial metric helpful in evaluating the contours of segmentations as well as the spatial positions of the voxels. The HD between two finite point sets A and B is defined as

$$\text{HD}(A, B) = \max(h(A, B), h(B, A))$$

$$h(A, B) = \|a - b\|$$

where a and b are all pixels within A and B.

In this challenge, we handled missing predictions with a DSC of 0, a HD of $+\infty$, and an ASD of $+\infty$.

Table 3. Overview of algorithms submitted to the PSFH Challenge.

| Team Name | Network | Loss Function | Post-processing | Patch Size | Optimizer | Initialization | Learning Rate | Cross-Validation | Epochs |
|---|---|---|---|---|---|---|---|---|---|
| angle_avengers | nnU-Net | Hausdorff distance and Focal | Ensemble learning | 256×256 | Stochastic Gradient Descent | Random | 1e-2 | Yes (5-fold) | 1,000 |
| Aloha | SAM | Cross-entropy and Dice | Ensemble learning | 512×512 | AdamW | Pre-trained SAM | 1e-4 | Yes (5-fold) | 400 |
| QiuYaoyang | UpNet | Dice | None | 256×256 | Adam | Random | 2e-2 | No | 40 |
| NKCGP | U-Net with Squeeze and Excitation Attention | Cross-entropy | Ensemble learning | 256×256 | AdamW | Random | 1e-3 | Yes (5-fold) | 50 |
| UCD_Med | U-Net with Transformer (TransU-Net) | Cross-entropy and Dice | None | 256×256 | AdamW | Pre-trained MiT-B0 | 1e-4 | No | 20 |
| Kunkunkk | U-Net with Transformer (Mix ViT) | Cross-entropy and Dice | None | 256×256 | Stochastic Gradient Descent | Pre-trained ResNet50 | 1e-2 | No | 150 |
| CQUT-Smart | U-Net with Transformer (BiFormer) | Dice | None | 256×256 | Adam | Pre-trained plain ViT | 1e-3 | No | 300 |
| aspirerabbit | U-Net | Dice | None | 256×256 | Adam | Random | 1e-4 | Yes (4-fold) | 200 |

Table 4. Overview of the data augmentation and pre-processing used in each algorithm.

| Team Name | Data Augmentation | External Dataset used | Pre-processing | Foundation model |
|---|---|---|---|---|
| angle_avengers | Default augmentation with stronger data augmentation (sDA5) | No | z-score normalization | No |
| Aloha | Horizontal flipping, Gaussian noise, blurring, random zooming, and affine transformation | Yes (pre-trained SAM) | up-sample the images to a resolution of 512×512 | Yes |
| QiuYaoyang | Random rotation, random resize, horizontal flip, Gaussian noise, Gaussian blur | No | normalization of the useful region | No |
| NKCGP | None | No | None | No |
| UCD_Med | Random rotation, vertically flip, horizontally flip, coarse Dropout | Yes (pre-trained MiT-B0) | normalization | No |
| Kunkunkk | Random rotation, vertically flip, horizontally flip | Yes (pre-trained ResNet50) | None | No |
| CQUT-Smart | Random rotation, vertically flip, horizontally flip | Yes (pre-trained plain ViT) | None | No |
| aspirerabbit | Random rotation, scaling, normalization | No | None | No |

2.5 Ranking

In order to ensure the fairness and reproducibility of algorithm comparison, we utilized the source codes (available at https://github.com/maskoffs/PS-FH-MICCAI23) provided by each team to calculate nine metrics for each case in Testing Set 2.

Firstly, box-and-whisker plots were employed to represent the distribution of metrics, highlighting the minimum and maximum values, the median, and the first (25%) and third (75%) quartiles, as well as identifying outliers.

Secondly, we employed five methods to create a ranking for each metric, including: 1) calculating the mean and then ranking the aggregated scores (MeanThenRank), 2) calculating the median and then ranking the aggregated scores (MedianThenRank), 3) calculating the ranking and then computing the mean of the aggregated ranks (RankThenMean), 4) calculating the ranking and then computing the median of the aggregated ranks (RankThenMedian), and 5) using the Wilcoxon signed-rank test to determine rankings based on the number of significant results (TestBased) (Wiesenfarth et al., 2021).

Thirdly, we applied RankThenMean specifically to the segmentation metrics to rank the segmentation performance of the different methods. Additionally, RankThenMean was applied to all metrics to determine the overall performance rankings. It is important to note that the results of the challenge were analyzed using the ChallengeR toolkit (available at https://github.com/wiesenfa/challengeR), which is specifically designed to calculate and display results for imaging challenges.

Finally, we investigated the stability of the final rankings. This involved implementing bootstrapping techniques to examine variations in the ranking positions of all teams. The ranking process was iteratively applied to 1000 bootstrap samples, each consisting of 700 images from Testing Set 2. The median Kendall's $\tau$ correlation coefficient between the rankings based on the full assessment dataset and the rankings for each bootstrap sample was used to assess the stability of the rankings across tasks (i.e., the nine metrics).

2.6 Further Analysis

**Model Design Predictors.** We conducted a comprehensive evaluation of the design choices implemented by the teams participating in the challenge. Our goal was to ascertain the influence of these design decisions on the teams' overall rankings and performance metrics. We assessed the statistical significance of performance variations across different sub-tasks by employing the Mann-Whitney U-test. This non-parametric test is particularly suited for comparing two independent samples that may not adhere to a normal distribution, providing robust insights into the differential impacts of various model design choices. We systematically analyzed the design choices, categorizing them into two principal groups:
Architectural Factors. We explored the impact of different network architectures on segmentation performance, focusing on several dimensions: ① Convolutional Neural Network (CNN) vs. Transformer-based models; ② U-Net-like structures vs. others; ③ Foundation models (e.g., SAM) vs. others; and ④ Pre-trained models vs. others.

Non-Architectural Factors. We analyzed the influence of non-architectural elements on performance enhancement, which are divided into four aspects: ① Preprocessing and data augmentation techniques (i.e., coarse dropout, flip and rotation, Gaussian noise application, non-zero normalization, scaling, and Z-score normalization); ② Loss functions (Hausdorff distance, Focal, Cross-entropy, and Dice losses); ③ Optimizers (Adam, AdamW, and Stochastic Gradient Descent); and ④ Postprocessing strategies (Ensemble vs. Non-ensemble approaches).

**Data Variability.** To evaluate the robustness of the algorithms, we generated various subsets of data to assess performance variations based on criteria such as data volume (e.g., Testing Set 1, Testing Set 2, and combined), data source or acquisition modality (e.g., SMU with Esaote My Lab vs. JNU with ObEye), and clinical impact measures (e.g., AoP≥120° vs. AoP<120°). We observed significant variability in rankings depending on the specific data subset examined, although 'Aloha' consistently maintained the top-ranking position across all categories.

**AoP Biometry Analysis.** Following the segmentation of the PS and FH regions, we performed ellipse fitting on each region. For the angle of progression (AoP) calculation, we defined two lines: the first being the long axis of the PS area, and the second extending tangentially from the right endpoint of the PS's long axis to the right contour of the FH area, as described by (Bai et al., 2022; Lu et al., 2022a). We calculated the AoP for each case in the Testing Set and compared these measurements to those determined by the challenge participants, using the absolute differences (ΔAoP) as a metric.

## 3. Experimental Results

We present a summary of the quantitative evaluation conducted on the methodologies implemented by all participating teams. Each method was assessed against the ground truth using two distinct testing datasets: Testing Set 1, which included 401 2D images, and Testing Set 2, comprising 700 2D images, as detailed in Sections 3.1 and 3.2, respectively. The results are conveyed through detailed statistical analysis. Box-and-whisker plots are used to illustrate the distribution of raw data points across the evaluations, highlighting the minimum and maximum values, the median, and the interquartile ranges (first quartile at 25% and third quartile at 75%), and identifying any statistical outliers. Throughout the manuscript, results are articulated as 'mean ± standard deviation.' This format was selected to provide a clear, concise representation of data distribution, facilitating straightforward interpretation of variations within the data. Such a presentation ensures that findings are accessible and interpretable, aiding in a comprehensive understanding of the effectiveness of the various computational methods employed.

3.1 Segmentation Performance on Testing Set 1

**Figure 3** provides a comprehensive visualization of the segmentation results achieved by the eight participating teams on Testing Set 1. It facilitates a comparison of segmentation accuracy using Dice Similarity Coefficient (DSC), Hausdorff Distance (HD), and Average Surface Distance (ASD) metrics across different anatomical structures.

**Figure 3A** details the DSC outcomes for three segmentation tasks: PSFH (top panel), FH (middle panel),

and PS (bottom panel). The DSC results display a range from 0.9184 to 0.9437 for PSFH, indicating consistently high performance across most teams. For FH, the range is tighter, from 0.9271 to 0.9498, demonstrating uniformly high precision among all methods. In contrast, PS segmentation exhibits a lower performance spectrum, with DSC values ranging from 0.8292 to 0.8902, reflecting variable accuracy across the teams. Notably, teams 'Aloha' and 'UCD_Med' achieved DSC scores above 0.8500 in PS segmentation, distinguishing themselves in this more challenging category.

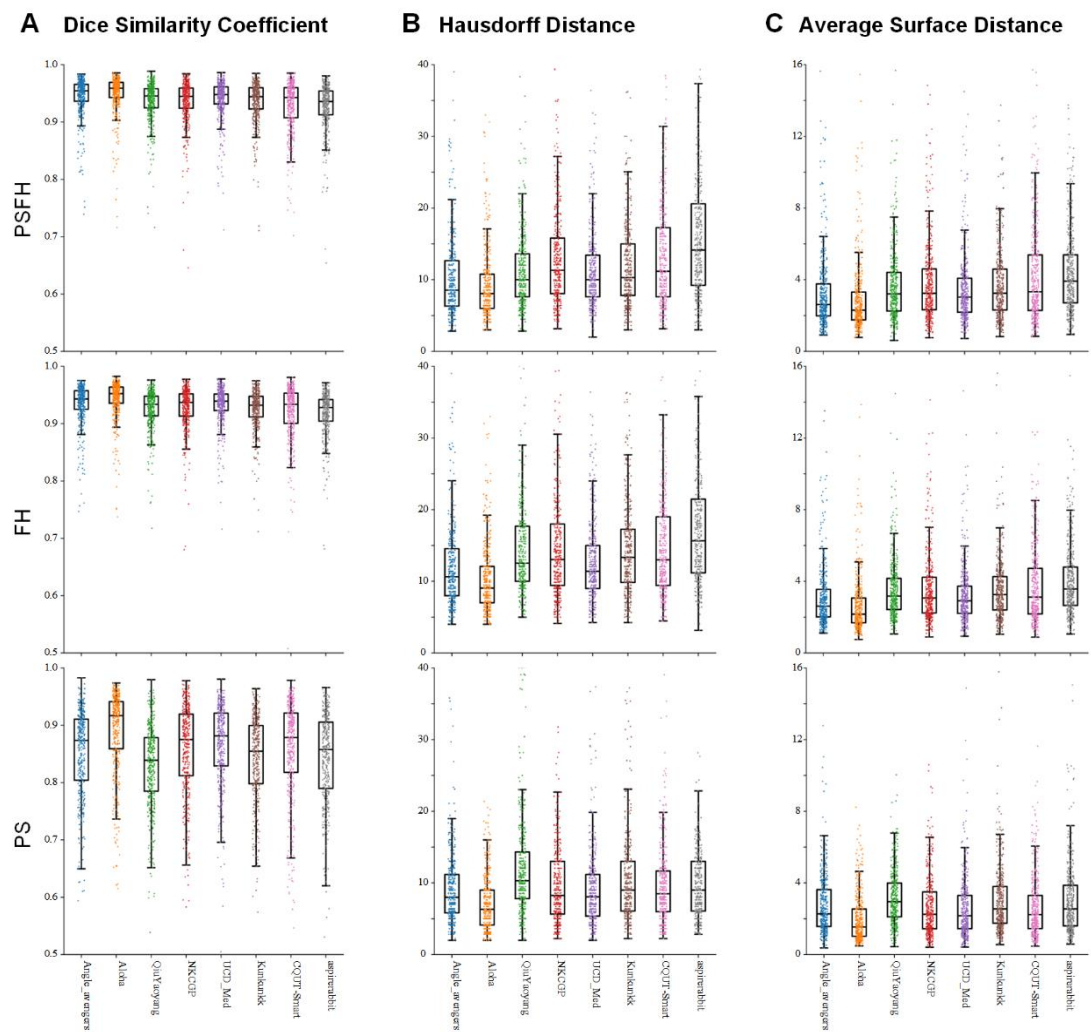

**Figure 3.** Segmentation results of 8 submitted methods on the testing set 1. PSFH, pubic symphysis and fetal head; FH, fetal head; PS, pubic symphysis.

**Figure 3B** presents the HD results, further elucidating the segmentation precision. For PSFH, HD values span from 10.380 to 19.967, while for FH, they range from 9.3896 to 17.813. The PS segmentation shows HD values between 7.0745 and 15.3202. Team 'Aloha' recorded the lowest HD in FH segmentation at 9.3896, indicating superior precision. Additionally, 'angle_avengers' and 'Aloha' achieved HD values below 10.0000 in PS segmentation, highlighting their exceptional accuracy.

**Figure 3C** showcases the ASD results, offering another perspective on segmentation effectiveness. The ASD ranges for PSFH are from 2.6747 to 4.0653, for FH from 2.9361 to 4.4663, and for PS from 2.0002 to 3.2702. 'Aloha' consistently performed well, maintaining an ASD below 3.0000 in FH segmentation. In the PS segmentation, 'angle_avengers', 'Aloha', and 'UCD_Med' reported ASD values below 2.800, demonstrating their commendable precision.

## 3.2 Competition on the Testing Set 2

To accurately reflect the performance of each team on Testing Set 2, we first analyzed the performance of each method on the segmentation evaluation metrics. Secondly, we analyzed the global and local rankings of each team across different tasks. Finally, we assessed the stability of these rankings.

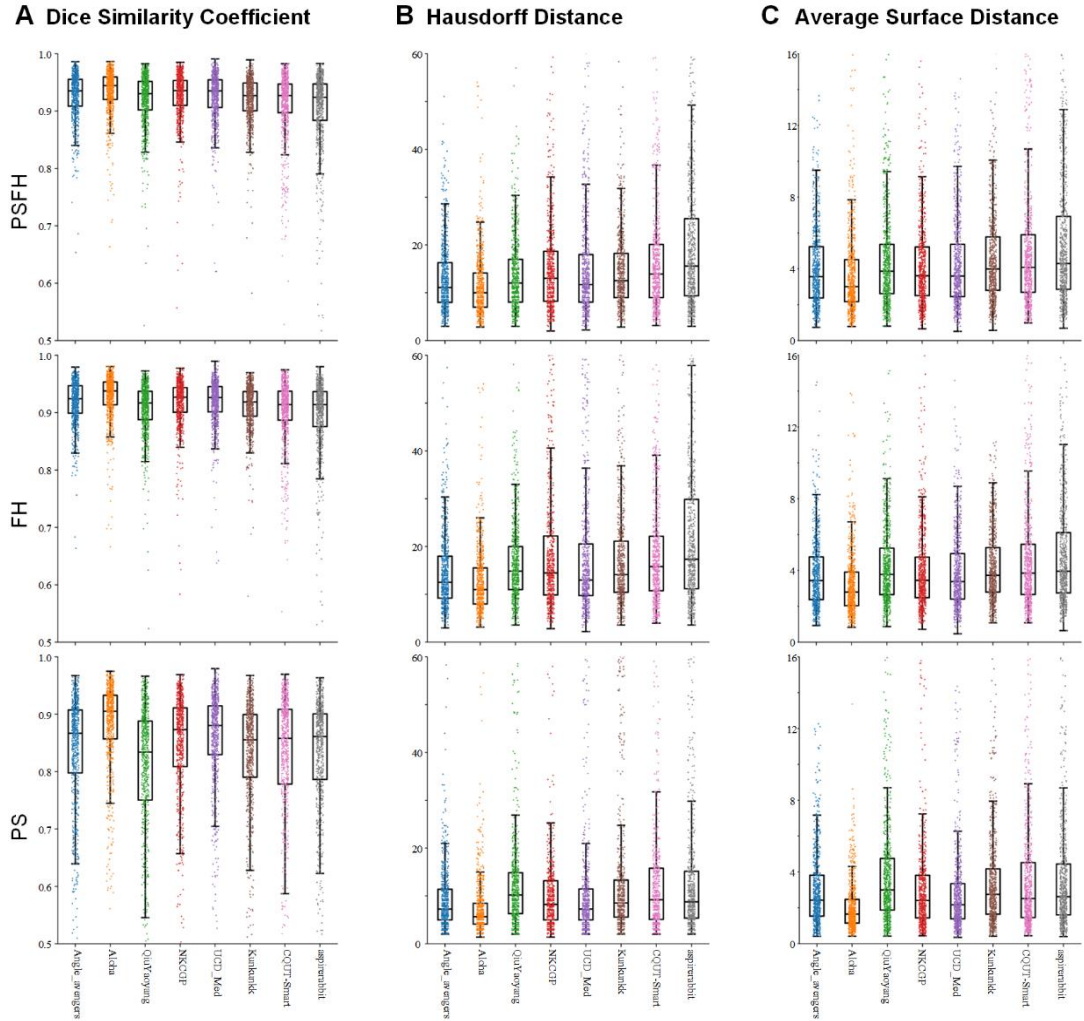

**Figure 4.** Segmentation results of 8 submitted methods on the testing set 2. The evaluation metrics include Dice Similarity Coefficient (A), Hausdorff Distance (B) and Average Surface Distance (C). PSFH, pubic symphysis and fetal head; FH, fetal head; PS, pubic symphysis.

**Quantitative Assessment. Figure 4** presents a comparison of segmentation accuracy using Dice Similarity Coefficient (DSC), Hausdorff Distance (HD), and Average Surface Distance (ASD) metrics across different anatomical structures. Figure 4A details the DSC outcomes for three segmentation tasks: PSFH (top panel), FH (middle panel), and PS (bottom panel). The DSC metric exhibited a range of 0.8949 to 0.9269 for PSFH, 0.9025 to 0.9327 for FH, and a significantly lower range of 0.7938 to 0.8717 for PS, with 'Aloha' achieving the highest DSC in both FH (0.9327) and PS (0.8717) segmentations. The second-highest DSC in FH segmentation was achieved by 'angle_avengers', and in PS segmentation by 'UCD_Med'. Figure 4B presents the HD results, further delineating segmentation quality, with recorded ranges of 13.2230 to 22.8544 for PSFH, 11.8598 to 20.2908 for FH, and 8.1839 to 19.0228 for PS. 'Aloha' continued to demonstrate exceptional precision in FH segmentation, achieving the lowest HD at 11.8598. Both 'angle_avengers' and 'Aloha' achieved an HD below 10.0000 in PS segmentation, underscoring their significant accuracy. Figure 4C showcases the ASD results, providing deeper insights into segmentation methodology effectiveness, with ranges from 3.3511 to 4.9997 for PSFH, 3.7634 to 5.6036 for FH, and 2.7117 to 4.3605 for PS. 'Aloha' achieved the lowest ASD in both FH (3.7634) and PS (2.7117) segmentations, with 'angle_avengers' and 'UCD_Med' following closely behind in their respective categories.

**Global and Local Ranking. Figure 5** delineates the global rankings of the eight participating teams across nine distinct metrics, highlighting the performance spread and relative consistency of each team. The global ranking is led by 'Aloha', followed by 'Angle_avengers', 'UCD_Med', 'NKCGP', 'Kunkunkk', 'QiuYaoyang', 'CQUT-Smart', and 'aspirerabbit'. This ranking illustrates a clear stratification of team performance, with top-performing teams securing higher ranks and those underperforming occupying the lower tiers. The ranking variability among the teams is relatively stable, with a maximum positional shift of three places observed, indicating a high confidence level in the ranking system. 'Aloha' dominates the global ranking, exhibiting a greater than 75% certainty of leading across all metrics, a testament to its superior performance and consistency. In contrast, teams ranked from second to sixth demonstrated variability and lower certainty levels for some metrics, suggesting competitive but inconsistent performances across different evaluation criteria. Teams in the middle of the ranking displayed moderate fluctuations, reflecting the competitive nature and the challenges in maintaining consistent performance across multiple metrics. The certainty of ranking decreases towards the lower end, with 'aspirerabbit' consistently positioned at the bottom with over 75% certainty in six out of the nine metrics evaluated. **Table 5** provides a detailed breakdown of the local rankings for each metric, where 'Aloha' consistently claims the top spot, underscoring its robust performance across different evaluation criteria. In the DSC/ASD metrics, 'UCD_Med', 'angle_avengers', and 'NKCGP' occupy the second to fourth positions, indicating their strong capabilities in achieving precise and accurate segmentations. The lower tier, comprised of 'Kunkunkk', 'QiuYaoyang', 'CQUT-Smart', and 'aspirerabbit', shows a significant performance disparity in these metrics. For the HD metrics, 'angle_avengers' consistently ranks second, while other teams show a maximum shift of 5 positions (i.e., 'QiuYaoyang').

Table 5. Local rankings (DSC, HD and ASD) for different anatomical structures (PSFH, FH and PS).

| Rank | DSC | | | HD | | | ASD | | |
|---|---|---|---|---|---|---|---|---|---|
| | PSFH | FH | PS | PSFH | FH | PS | PSFH | FH | PS |
| 1 | Aloha | Aloha | Aloha | Aloha | Aloha | Aloha | Aloha | Aloha | Aloha |
| 2 | UCD_Med | angle_avengers | UCD_Med | angle_avengers | angle_avengers | angle_avengers | UCD_Med | NKCGP | UCD_Med |
| 3 | NKCGP | NKCGP | NKCGP | UCD_Med | QiuYaoyang | UCD_Med | angle_avengers | angle_avengers | NKCGP |
| 4 | angle_avengers | UCD_Med | angle_avengers | Kunkunkk | UCD_Med | NKCGP | NKCGP | UCD_Med | angle_avengers |
| 5 | Kunkunkk | QiuYaoyang | CQUT-Smart | NKCGP | NKCGP | Kunkunkk | Kunkunkk | QiuYaoyang | CQUT-Smart |
| 6 | QiuYaoyang | Kunkunkk | aspirerabbit | QiuYaoyang | Kunkunkk | CQUT-Smart | QiuYaoyang | Kunkunkk | Kunkunkk |
| 7 | CQUT-Smart | CQUT-Smart | Kunkunkk | CQUT-Smart | CQUT-Smart | aspirerabbit | CQUT-Smart | CQUT-Smart | aspirerabbit |
| 8 | aspirerabbit | aspirerabbit | QiuYaoyang | aspirerabbit | aspirerabbit | QiuYaoyang | aspirerabbit | aspirerabbit | QiuYaoyang |

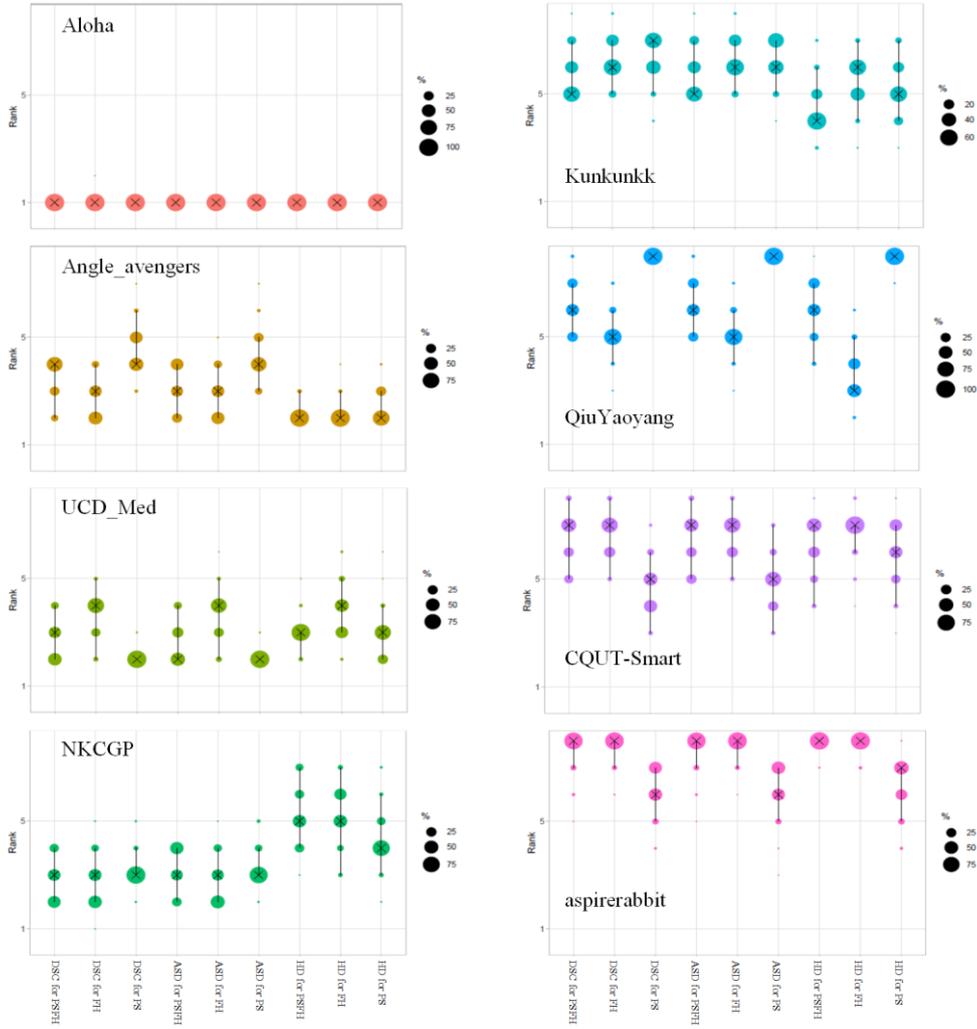

**Figure 5.** Blob plots for visualizing ranking stability of each algorithm ('Aloha', 'Angle_avengers', 'UCD_Med', 'NKCGP', 'Kunkunkk', 'QiuYaoyang', 'CQUT-Smart', or 'aspirerabbit') across each metric/task (DSC for PSFH, DSC for FH, DSC for PS, ASD for PSFH, ASD for FH, ASD for PS, HD for PSFH, HD for FH, or HD for PS). The area of each blob at position (A, rank j) is proportional to the relative frequency A achieved rank j (here across b=1000 bootstrap samples). The median rank for each algorithm is indicated by a black cross. 95% bootstrap intervals across bootstrap samples (ranging from the 2.5th to the 97.5th percentile of the bootstrap distribution) are indicated by black lines.

**Ranking Stability.** To determine the significance of a team outperforming another in terms of individual metrics, we employed RankThenMean compared with other approaches, and used the Wilcoxon signed-rank test with Holm's adjustment for multiple testing for each metric. The challenge winner 'Aloha' secured the top position for all metrics, and the team 'aspirerabbit' at the bottom of the rankings for most metrics are relatively stable. For DSC (**Figure 6**), HD (**Figure 7**) and ASD (**Figure 8**), 'Aloha' ranks first in segmenting all targets, while 'aspirerabbit' ranks last in segmenting PSFH and FH, and 'QiuYaoyang' ranks last in segmenting PS. High-ranking teams robustly outperform lower-ranked teams, and statistical significances were observed when comparing all image metrics between a team and another team ranked for each metric. However, transitioning from RankThenMean to the four other ranking approaches caused substantial changes for the middle-ranked teams. When changing from RankThenMean to MeanThenRank, MedianThenRank, RankThenMedian, and TestBased, respectively, maximum shifts are 5 for 'NKCGP', 2 for 'Angle_avengers' and 'NKCGP', 3 for 'CQUT-Smart', and 2 for 'NKCGP', 'aspirerabbit', 'QiuYaoyang', 'UCD_Med' and 'NKCGP'. Despite these variations, the distribution of Kendall's tau between the ranking based on the full assessment dataset and the ranking for each bootstrap sample illustrates that RankThenMean can provide very stable ranking of all algorithms (**Figure 9**).

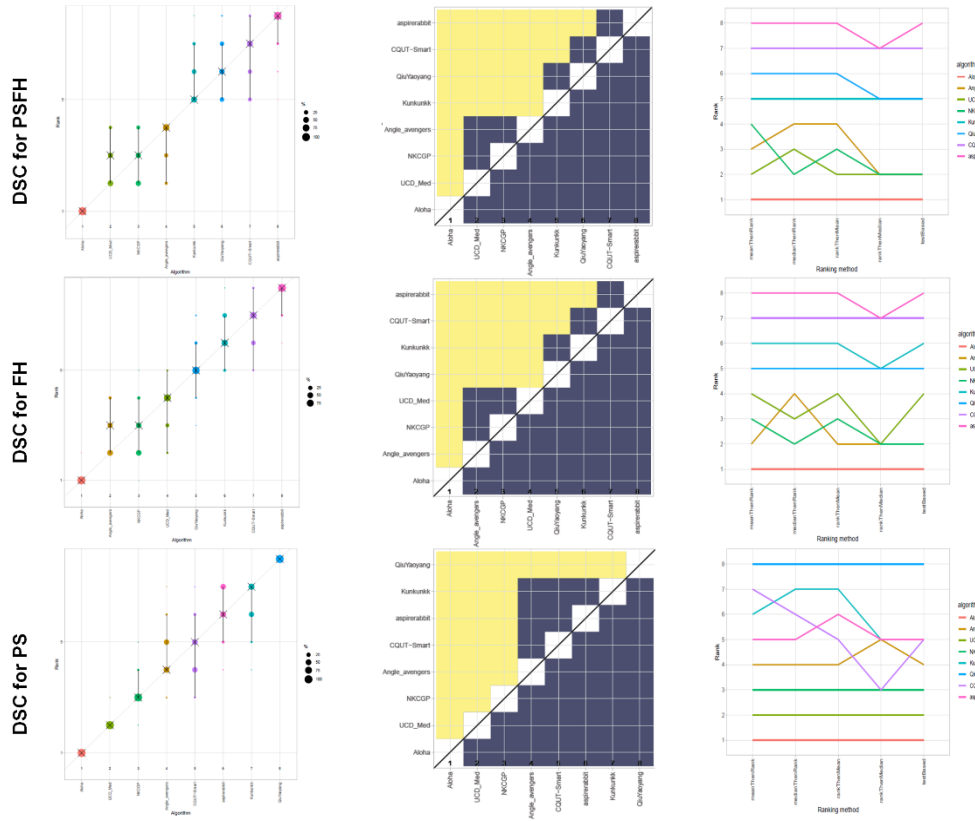

**Figure 6.** Ranking of eight algorithms ('Aloha', 'Angle_avengers', 'UCD_Med', 'NKCGP', 'Kunkunkk', 'QiuYaoyang', 'CQUT-Smart', and 'aspirerabbit') across three metrics/tasks (DSC for PSFH, DSC for FH, and DSC for PS). Left panel: Blob plots for visualizing ranking stability based on bootstrap sampling. Middle panel: Ranking heatmaps for visualizing assessment data. Each cell (i, algorithm) shows absolute frequency of test cases in which algorithm achieved rank i. Right panel: Line plots for visualizing the robustness of ranking across different ranking methods. Each algorithm is represented by one colored line. For each ranking method (MeanThenRank, MedianThenRank, RankThenMean, RankThenMedian or TestBased) encoded on the x-axis, the height of the line represents the corresponding rank. Horizontal lines indicate identical ranks for all methods.

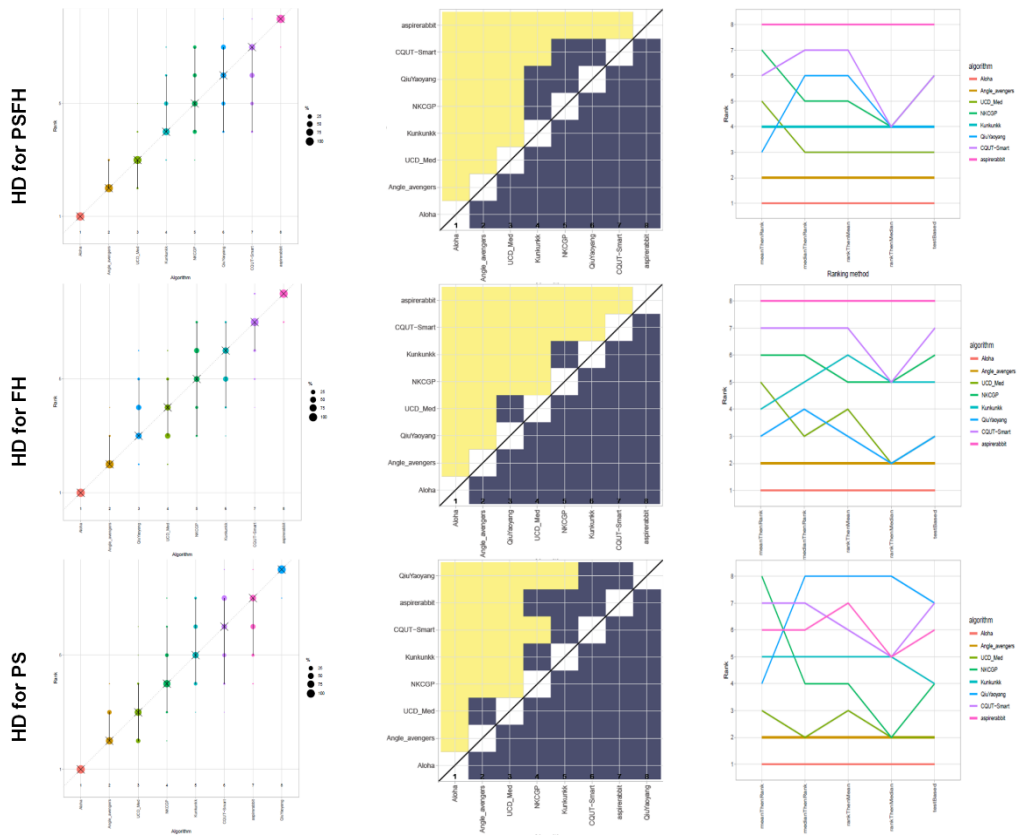

**Figure 7.** Ranking of eight algorithms ('Aloha', 'Angle_avengers', 'UCD_Med', 'NKCGP', 'Kunkunkk', 'QiuYaoyang', 'CQUT-Smart', and 'aspirerabbit') across three metrics/tasks (HD for PSFH, HD for FH, or HD for PS). Left panel: Blob plots for visualizing ranking stability based on bootstrap sampling. Middle panel: Ranking heatmaps for visualizing assessment data. Each cell (i, algorithm) shows absolute frequency of test cases in which algorithm achieved rank i. Right panel: Line plots for visualizing the robustness of ranking across different ranking methods. Each algorithm is represented by one colored line. For each ranking method (MeanThenRank, MedianThenRank, RankThenMean, RankThenMedian or TestBased) encoded on the x-axis, the height of the line represents the corresponding rank. Horizontal lines indicate identical ranks for all methods.

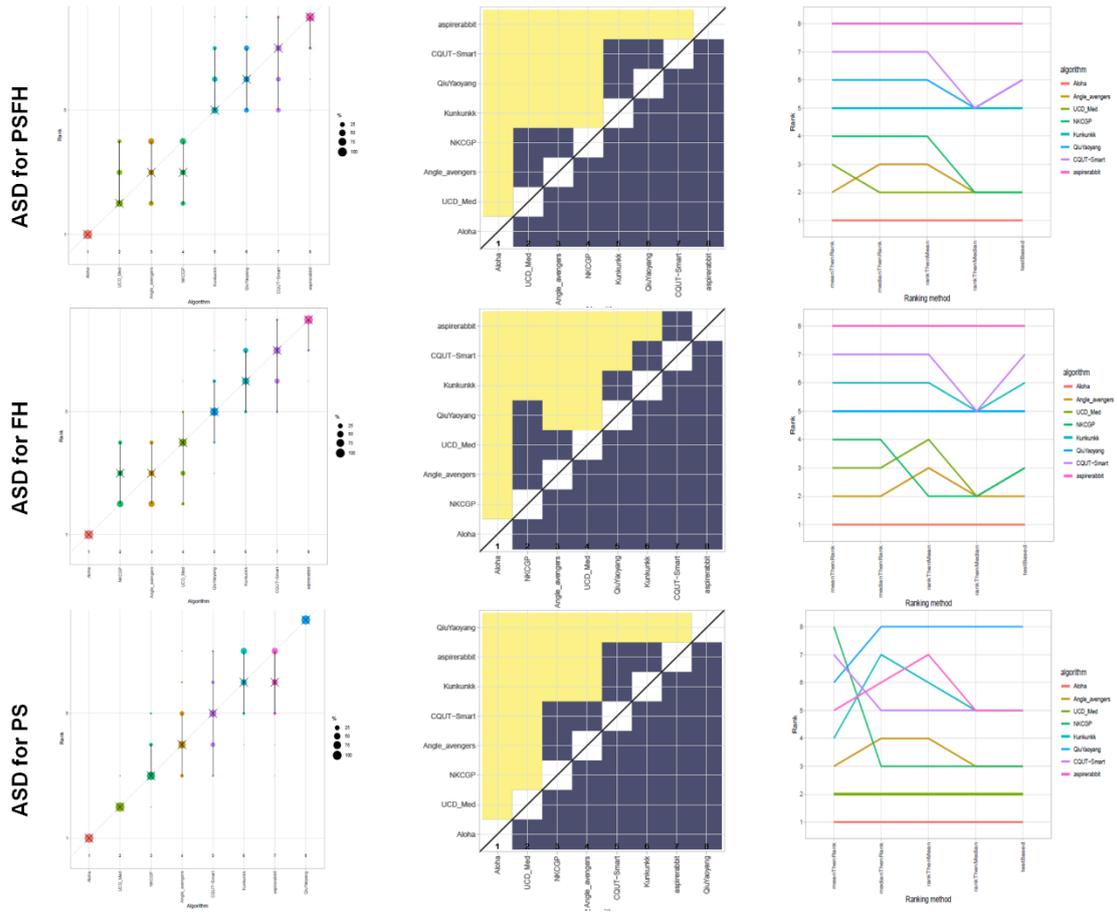

**Figure 8.** Ranking of eight algorithms ('Aloha', 'Angle_avengers', 'UCD_Med', 'NKCGP', 'Kunkunkk', 'QiuYaoyang', 'CQUT-Smart', and 'aspirerabbit') across three metrics/tasks (ASD for PSFH, ASD for FH, and ASD for PS). Left panel: Blob plots for visualizing ranking stability based on bootstrap sampling. Middle panel: Signifcance maps for visualizing ranking stability based on statistical signifcance. Tey depict incidence matrices of pairwise signifcant test results e.g. for the one-sided Wilcoxon signed rank test at 5% signifcance level with adjustment for multiple testing according to Holm. Yellow shading indicates that metric values of the algorithm on the x-axis are signifcantly superior to those from the algorithm on the y-axis, blue color indicates no signifcant superiority. Right panel: Line plots for visualizing the robustness of ranking across different ranking methods. Each algorithm is represented by one colored line. For each ranking method (MeanThenRank, MedianThenRank, RankThenMean, RankThenMedian or TestBased) encoded on the x-axis, the height of the line represents the corresponding rank. Horizontal lines indicate identical ranks for all methods.

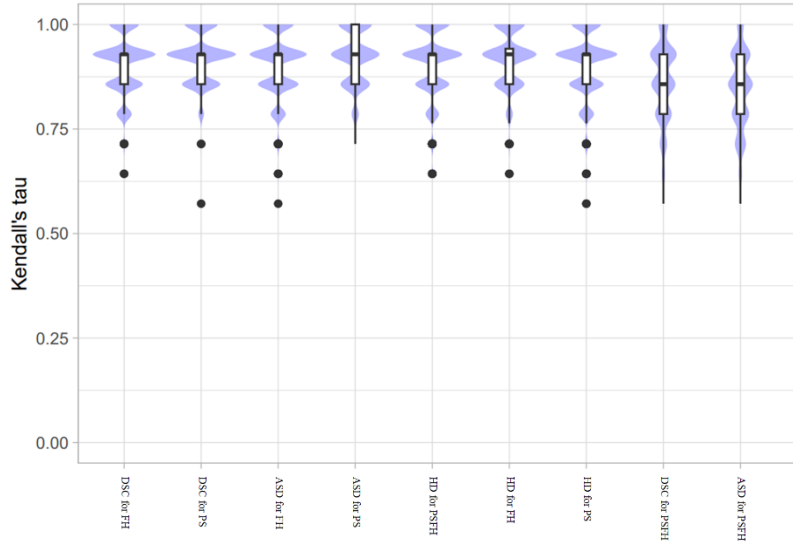

**Figure 9.** Violin plots for visualizing ranking stability based on bootstrapping. The ranking list based on the full assessment data is compared pairwise with the ranking lists based on the individual bootstrap samples (here b = 1000 samples). Kendall's tau is computed for each pair of rankings, and a violin plot that simultaneously depicts a boxplot and a density plot is generated from the results.

## 4. Discussion

To our knowledge, this study is the most comprehensive effort to date to systematically benchmark a variety of approaches submitted to a global challenge focused on PSFHS from 2D intrapartum ultrasound images. Developed using the largest and highest-quality dataset of 2D intrapartum ultrasound images and annotations currently available, our 2023 PSFHS Challenge has set a new standard in the field. The benchmarking study encompasses methodologies developed by 8 groups, representing a well-curated subset of the 193 participants. While there are previous efforts in this area (Bai et al., 2022; Chen et al., 2024; Lu et al., 2022a), our study significantly contributes in several ways. Firstly, our study is among the few that focus on the segmentation of both fetal and maternal structures within 2D intrapartum ultrasound images—an endeavor that is inherently more challenging due to poor quality of ultrasound images caused by multiple factors such as fetal movement, posture, and position. Secondly, this challenge attracted more participants than any prior study and provided a dataset that is not only publicly available but also multi-center and multi-device in nature, offering a scale that far surpasses previous challenges (Rueda et al., 2014; van den Heuvel et al., 2018). Moreover, the segmentation task in our study is substantially more complex than those in established prior challenges, which typically focused on single fetal structures. Our challenge involves the class-imbalanced segmentation of multiple targets— simultaneously segmenting the fetal skull and maternal pelvis—making it a uniquely difficult endeavor. Furthermore, the proposed methods have demonstrated superior robustness and effectiveness, as evidenced by the diversity of approaches and the range of performance metrics evaluated. Lastly, our study delves deeply into the key factors that optimize segmentation performance through extensive post-challenge analyses of the 8 submitted approaches, providing valuable insights into the nuances of methodological advancements in this area.

## 4.1 Results on the Whole Testing Set

When we combine results from Testing Set 1 and Testing Set 2, **Figure 10** provides a comparative analysis of segmentation performance on DSC, HD, and ASD metrics. Specifically, the DSC metric exhibited a range of 0.9034 to 0.9330 for PSFH, 0.9114 to 0.9389 for FH, and 0.8261 to 0.8785 for PS. The HD metric ranges from 12.1874 to 21.8029 for PSFH, 10.9601 to 19.3884 for FH, and 7.7799 to 17.6743 for PS. The recorded ASD ranges from 3.1048 to 4.6594 for PSFH, 3.4621 to 5.1894 for FH, and 2.4526 to 3.9286 for PS. Overall, team 'Aloha' achieves the highest DSC and the lowest HD and ASD across the metrics. We also compared the results achieved by all teams on Testing Set 1 with those on Testing Set 2. The average metrics on Testing Set 1 are as follows: DSC for PS is 0.8526, DSC for FH is 0.9371, DSC for PSFH is 0.9286, HD for PS is 11.1666, HD for FH is 12.5926, HD for PSFH is 14.5130, ASD for PS is 2.8519, ASD for FH is 3.7260, and ASD for PSFH is 3.4430. Conversely, the average metrics on Testing Set 2 are: DSC for PS is 0.8298, DSC for FH is 0.9199, DSC for PSFH is 0.9114, HD for PS is 13.8598, HD for FH is 15.8062, HD for PSFH is 18.1934, ASD for PS is 3.7068, ASD for FH is 4.6146, and ASD for PSFH is 4.2158. This observation suggests that the segmentation methods of all participating teams performed better on Testing Set 1 than on Testing Set 2, likely due to the design of our challenge and the distribution of training and testing datasets. Specifically, our training data includes 3743 2D ultrasound images from Nanfang Hospital of SMU and 257 from Zhujiang Hospital, while Testing Set 1 comprises 301 images from Zhujiang Hospital of SMU and 100 from the First Affiliated Hospital of JNU. In contrast, Testing Set 2 includes 487 images from Zhujiang Hospital of SMU and 213 from the First Affiliated Hospital of JNU. Additionally, the distribution range of AoP corresponding to the training data set spans 60° to 180°, whereas the AoP distribution for Testing Set 1 ranges from 50° to 160°, and for Testing Set 2, it extends from 50° to 180°.

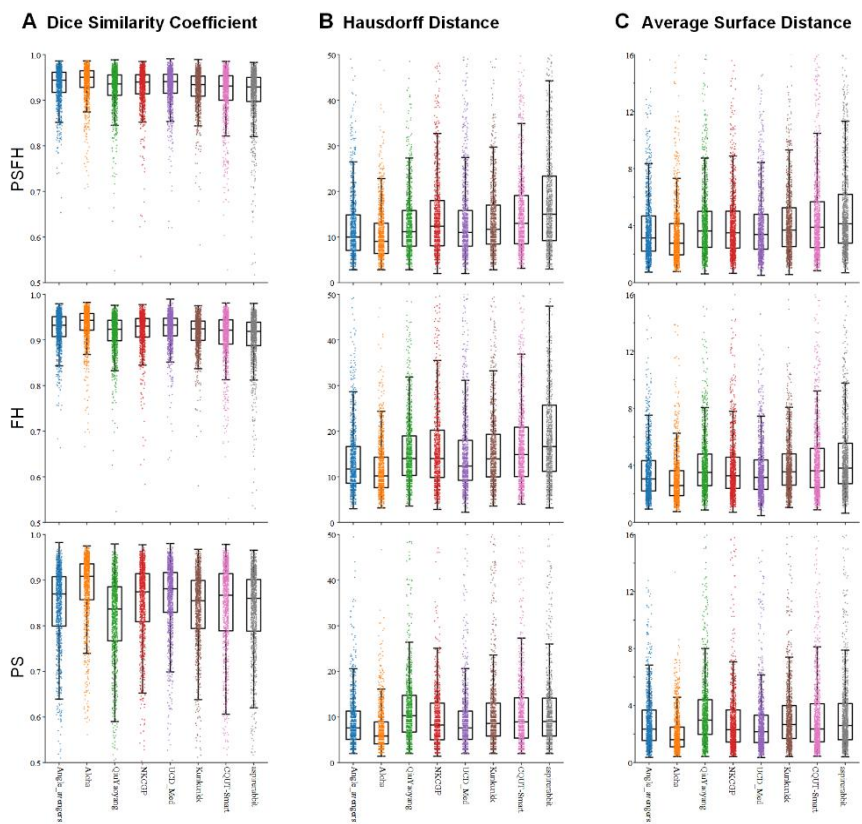

**Figure 10.** Segmentation results of 8 submitted methods on the whole testing set. PSFH, pubic symphysis and fetal head; FH, fetal head; PS, pubic symphysis.

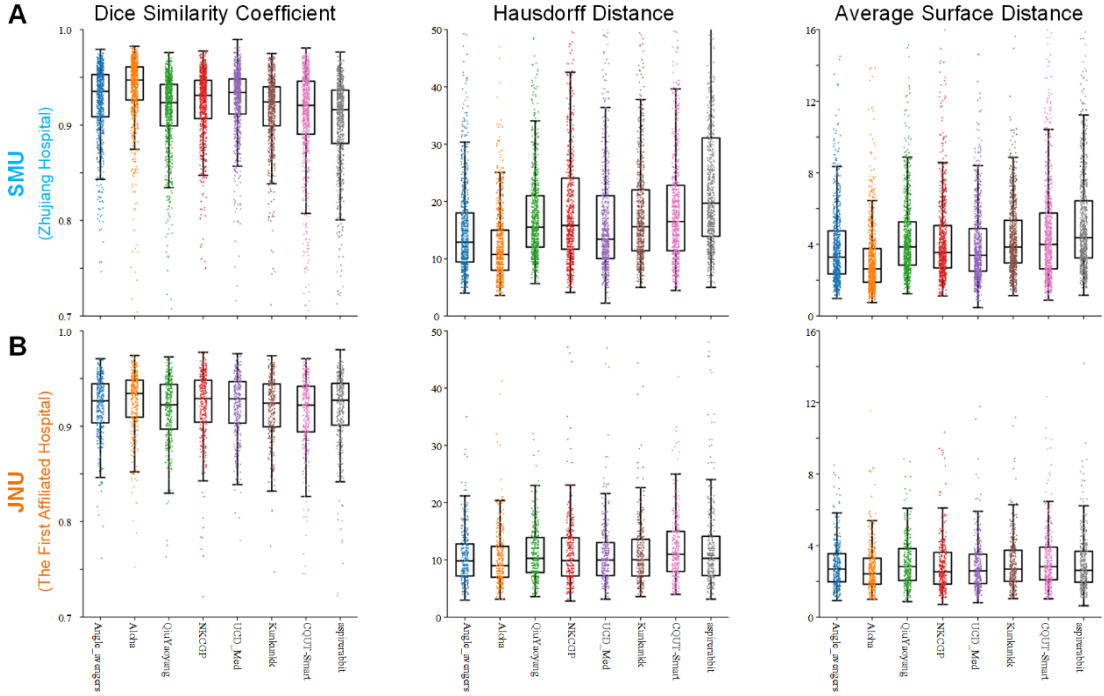

**Figure 11.** The PSFH segmentation performance of 8 submitted methods on the testing subsets from two different sources (i.e., SMU (Zhujiang Hospital) **(A)** and JNU (The First Affiliated Hospital) **(B)**).

The entire test dataset comprises two parts: one part from JNU using ultrasound equipment 'ObEye' and the other from SMU using 'Esaote My Lab'. We analyzed the segmentation performance of eight methods on these datasets. **Figure 11** presents a comparative analysis of PSFH segmentation performance on DSC, HD, and ASD. The metric ranges on the SMU set are 0.8977 to 0.9363 for DSC, 12.9248 to 25.5377 for HD, and 3.2424 to 5.2928 for ASD, whereas on the JNU set, the ranges are 0.9178 to 0.9363 for DSC, 10.3308 to 12.4994 for HD, and 2.7583 to 3.2512 for ASD. This analysis indicates that the segmentation methods generally perform better on the JNU set than on the SMU set, possibly due to differences in the training data distribution and ultrasound equipment used. **Figure 12** shows the visual comparisons of the segmentation results obtained by all 8 submitted algorithms on cases of the JNU and SUM sets.

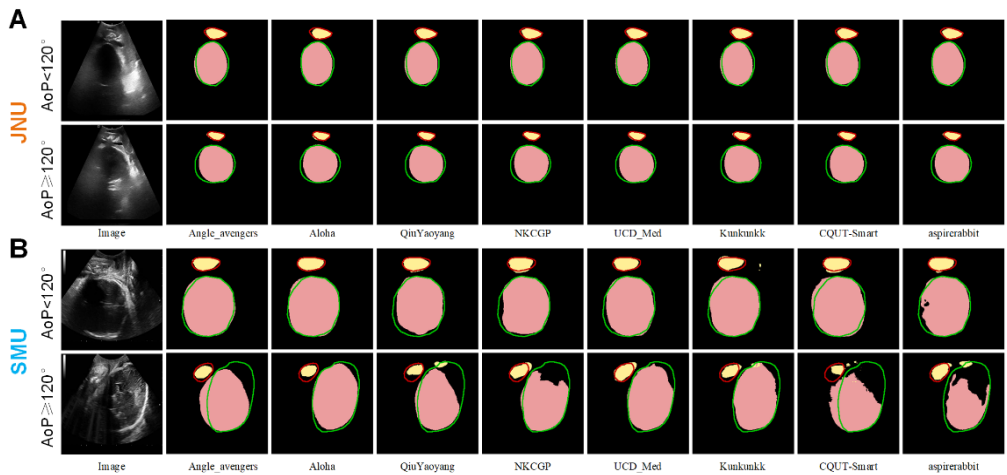

**Figure 12.** Visual comparison of segmentation results of the 8 submitted methods on the testing data. **A)** Samples (i.e., AoP<120° and AoP≥120°) collected from SMU. **B)** Samples (i.e., AoP<120° and AoP≥120°) collected from JNU. The ground truths

represented by contours and predictions of pubic symphysis (PS) and fetal head (FH) are displayed in different colors.

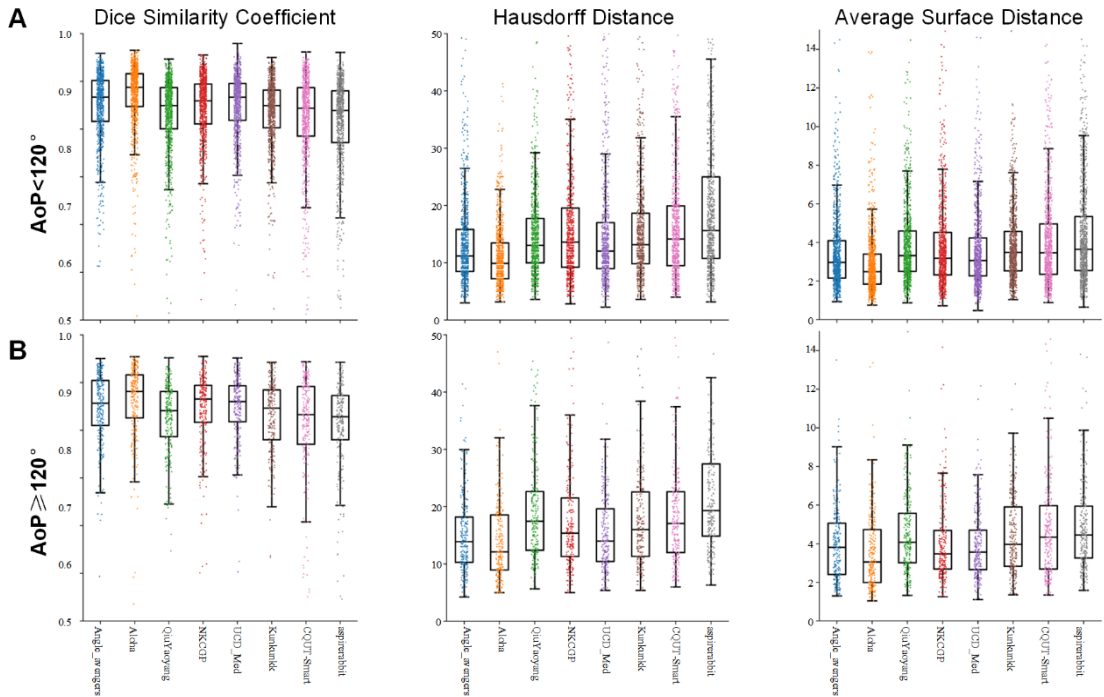

**Figure 13.** The PSFH segmentation performance of 8 submitted methods on the two subsets (i.e., AoP<120° (**A**) and AoP≥120° (**B**)) of the whole testing dataset (set 1 and 2).

Further, we divided the entire testing set into two parts based on AoP: AoP<120° and AoP≥120° sets. **Figure 13** compiles a comparative analysis of the PSFH segmentation performance across these sets, with metric ranges on the AoP<120° set being 0.9042 to 0.9341 for DSC, 11.4137 to 21.4999 for HD, and 2.9422 to 4.5668 for ASD. For the AoP≥120° set, the ranges are 0.9006 to 0.9290 for DSC, 14.9928 to 22.9014 for HD, and 3.6945 to 4.9952 for ASD. These results suggest that overall, the segmentation methods of all participating teams perform similarly on both AoP subsets. This finding is likely influenced by our challenge design, where the training data contains 3577 images with AoP<120° and 423 with AoP≥120°, while the testing data contains 723 images with AoP<120° and 378 with AoP≥120°.

4.2 Characteristics of Top-Performing Model

We explored the effects of different network architectures on segmentation performance. CNNs were used by three teams. A pure transformer model was employed by the 'Aloha' team, which substantially outperformed the other teams using CNNs (N=3) or hybrid models combining CNN and Transformer technologies (N=4) (**Figure 14A**). Most models adopted U-Net-like architectures (N=5), which were found to be less effective than the other two architectures, namely UperNet and SAM (**Figure 14B**). Out of the eight teams, two used complex methods: nnU-Net covers the entire pipeline from preprocessing to model configuration, training, postprocessing, and ensembling; SAM is a foundational model capable of segmenting any image using zero-shot learning and generalizing across domains without additional training. It was observed that the team using SAM performed better than the team employing nnUNet (**Figure 14C**). Four teams used pre-trained models and implemented further enhancements to the original architecture in an effort to improve segmentation performance (**Figure 14D**).

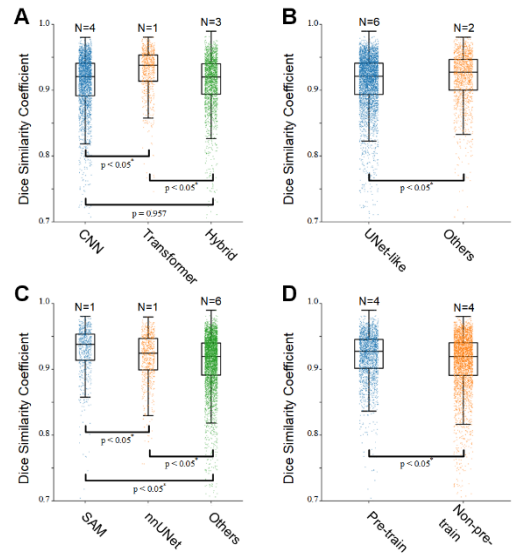

**Figure 14.** Comparative summaries of the PSFH segmentation performance of the 8 submitted methods. **A)** One Transformer-based method outperformed the four CNN methods and three hybrid methodologies with statistical significance. **B)** Two non-UNet-like methods outperformed the six UNet-like methods with statistical significance. **C)** Both of Segment Anything Model (SAM) and nnUNet notably exceeded the performance of the other six contenders. Importantly, SAM achieved a statistically significant lead over nnUNet. **D)** Four solutions utilizing pre-trained models surpassed these methods without pre-trained elements with statistical significance.

We also investigated the effects of non-architectural elements on performance enhancement. All teams performed data preprocessing or data augmentation, which to some extent improved segmentation accuracy (**Figure 15A**). Five different loss functions were employed by the eight teams; however, relying solely on the Dice loss function did not yield optimal segmentation results (**Figure 15B**). Six of the eight teams used the AdamW/Adam optimizer, achieving better results than those using the SGD optimizer (**Figure 15C**). The top four teams employed an ensemble learning method to enhance segmentation accuracy (**Figure 15D**).

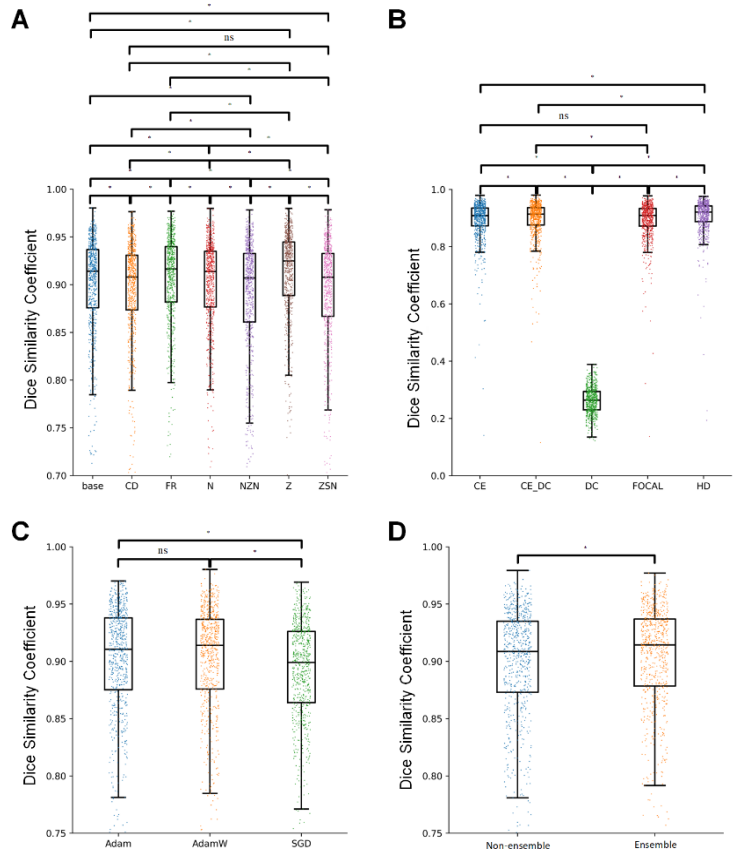

**Figure 15.** Comparative summaries of the PSFH segmentation performance. **A)** Effects of different methods of data augmentation on segmentation performance. **B**) Effects of different loss functions on segmentation performance. **C**) Effects of different optimizers on segmentation performance. **D**) Effects of post-processing methods on segmentation performance. base - without data augmentation, CD - coarse dropout, FR - flip and rotation, N - Gaussian noise application, NZN - non-zero normalization, Z - Z-score normalization, ZSN - scaling and Z-score normalization, CE - Cross-entropy loss, CE_DC - Cross-entropy and Dice loss, DC – Dice loss, Focal – Focal loss, HD - Hausdorff distance loss, SGD - Stochastic Gradient Descent.

The winner, 'Aloha', utilized all network architectural and non-network architectural factors known to improve segmentation performance. Specifically, they were the only team to upscale images from 256×256 to 512×512 to ensure compatibility with the transformer-based foundational model, SAM. They enriched the training dataset using diverse data augmentation techniques from the MONAI library, including horizontal flipping, Gaussian noise, blurring, random zooming, and affine transformations. They incorporated a warm-up phase and used the AdamW optimizer to adjust a weighted combination of cross-entropy and Dice losses. Their methodology involved alternating between five sets of trained LoRA parameters to produce ten predictions per image, which were then combined using a soft ensemble method to reduce variance and enhance result robustness. The final post-processing step focused on isolating the largest connected component from each detected object. However, the enhanced performance of SAM is accompanied by an increase in model complexity, reflected in a significant uptick in parameter count, which poses challenges during training and deployment. Given the lack of medical resources and expertise in underdeveloped areas, we also need to further optimize the SAM so that it can be implemented on cost-effective mobile platforms, thereby aiding healthcare professionals in enhancing local medical practices.

4.3 Clinical Impact

Transperineal ultrasound has been recognized as an effective tool for determining fetal head station during labor. The Angle of Progression (AoP) is a reliable parameter to assess fetal head station during labor. Research on the relationship between AoP and fetal head station indicates that the corresponding AoP for fetal station ranging from -3.0 to 5.0 spans from 84° to 170°. In our study, the AoP range for the training dataset was 60° to 180°, for training dataset 1 it was 50° to 160°, and for training dataset 2 it was 50° to 180°. The AoP for each image in the testing sets was calculated using all labels (excluding the background) and compared to the AoP determined by each participant in the challenge. The differences in AoP (ΔAoP) across testing sets 1 and 2 and the combined set (1+2) ranged from 5.6890-8.7844°, 8.2620-10.7620°, and 7.3250-10.0545°, respectively (**Figure 16**). These observations indicate that the average ΔAoP of all participating teams is less than 11°, corresponding to a change in fetal station of approximately 1.0 (Tutschek et al., 2013).

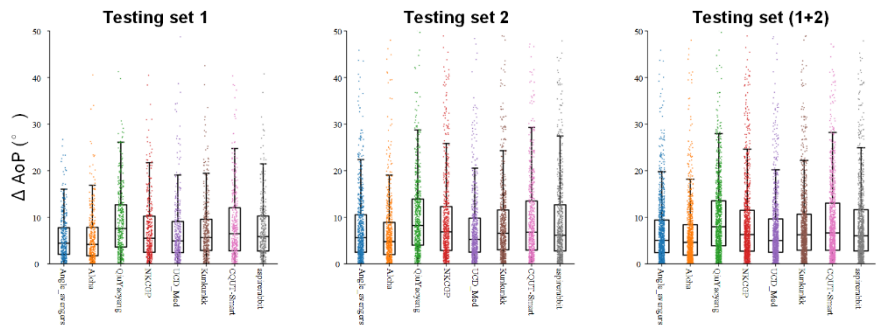

**Figure 16.** The AoP measurement performance of 8 submitted methods on the testing sets.

4.4 Limitations of the PSFHS challenge

The dataset utilized in this study encompasses a broad spectrum of fetal positions, albeit with notable imbalances in the distribution of data across different AoPs and variations in image acquisition across hospitals and devices. Specifically, the dataset exhibits a significant disparity in the number of images with AoP less than 120° (4594 images) compared to those with AoP equal to or greater than 120° (507 images), despite the observed similarity in performance metrics. This imbalance is crucial for clinical decision-making, particularly in predicting delivery mode, and underscores the need for careful consideration in model development and validation.

Furthermore, the dataset's geographical and methodological homogeneity, limited to Guangdong Province of China and two specific types of equipment, restricts the generalizability of the models developed. The exclusion of data from diverse regions and countries, along with variations in imaging protocols and operator experience, limits the applicability of the findings to a narrow context(Zhou et al., 2023). This limitation is particularly pertinent in the context of global healthcare disparities and the need for models that can be adapted to different clinical settings and imaging conditions (Ghi et al., 2022; Ramirez Zegarra et al., 2024)(Vesal et al., 2022).

The variability in image quality, attributed to the diverse operator experience, is a critical factor that could influence the segmentation algorithms' performance. Despite this, the study's findings suggest that both nnUNet and SAM demonstrated robust performance across the dataset. This observation, while encouraging, underscores the need for further investigation into the impact of image quality on algorithm performance and the development of strategies to mitigate these effects.

The further study's focus on the development of models that are computationally efficient and scalable, with an eye towards implementation on cost-effective mobile platforms, reflects a commendable effort to address global healthcare challenges. However, the goal of refining SAM through knowledge distillation to enhance its applicability on less sophisticated hardware is a significant undertaking that requires careful consideration of the trade-offs between model performance and computational requirements.

The primary clinical objective of the study, which involves segmenting targets, fitting ellipses to these segments, and calculating the AoP, presents a unique challenge in the accuracy of AoP estimation, primarily due to the ellipse fitting step. The potential for errors in landmark determination and subsequent AoP calculation due to ellipse fitting inaccuracies highlights the need for further refinement in the segmentation and fitting algorithms to improve the overall accuracy of the AoP estimation process.

Lastly, the absence of data on the mother's final mode of delivery precludes the prediction of delivery mode solely from ultrasound images and parameters, which is a limitation that could be addressed through future studies that incorporate this critical clinical outcome.

4.5 Future direction

PSFHS2023 has set out to advance the state-of-the-art in PS and FH segmentation for labor progression assessments. While the results are promising, these tasks have not yet been solved during this challenge. The dataset only provided 2D static images containing FH and PS (i.e., standard plane). Labor assessment via ultrasound videos maybe more challenging (Cai et al., 2020; Qiu et al., 2024). The positive reception to PSFHS2023 has spurred the development of the Intrapartum Ultrasound Grand Challenge (IUGC) 2024 of MICCAI 2024 (https://codalab.lisn.upsaclay.fr/competitions/18413) which aims to expand the challenge beyond static 2D images to more unexplored regions of ultrasound video close to clinical applications (Alsharid et al., 2022; Li et al., 2021; Sharma et al., 2021). Furthermore, addressing the limitations in data preparation and two-stage biometry discussed earlier will enhance the analysis of future challenges.

Lastly, we have released the whole testing set and the source code of this challenge teams. This offer opportunities to validate and enhance the statistical robustness of the challenge's conclusions, enabling other researchers to compare their methods with the results of PSFHS2023.

## 5. Conclusion

While intrapartum ultrasound-based labor progress assessments have become a clinical reality, the PSFHS Challenge represents a key advance in automated intrapartum ultrasound image analysis for labor assessment. This challenge marks the first multicenter challenge with a substantial dataset, serving as a catalyst for further innovation in intrapartum care. Participants demonstrated their ability to segment PS and FH, demonstrating high DICE and low HD and ASD. These achievements highlight the potential of deep learning for intrapartum ultrasound image segmentation. However, it is important to recognize that the clinical applicability of automated labor assessment may not be fully captured by segmentation metrics alone. Nonetheless, these significant advances hold the promise of reducing reliance on traditional manual image analysis and increasing the efficiency of labor assessment.

## Code availability

Evaluation code used within the challenge, along with example notebooks can be found at the following repository: https://github.com/lanesra10/FH-PS-AOP-grandchallenge. Code used for running the eight methods is available at https://github.com/maskoffs/PS-FH-MICCAI23).

## Ethics approval

The data collected for this challenge received approval from the institutional review boards of Zhujiang Hospital of Southern Medical University (No. 2023-SYJS-023), Nanfang Hospital of Southern Medical University (No. NFCE-2019–024) and the First Affiliated hospital of Jinan University (No. JNUKY-2022-019). Informed consent was waived because of the retrospective nature of the study and the analysis used anonymous medical image data.

## Authors contribution

Organizers

**Jieyun Bai:** Conceptualization, Formal Analysis, Investigation, Methodology, Software, Validation, Visualization, Writing– original draft, Writing– review & editing. **Zihao Zhou**: Conceptualization, Formal Analysis, Investigation, Methodology, Software, Validation, Visualization, Writing original draft, Writing– review & editing. **Zhanhong Ou**: Conceptualization, Data curation, Formal Analy sis, Investigation, Resources, Visualization, Writing– review & editing. **Dong Ni**: Conceptualization, Funding acquisition, Writing– review & editing. **Mei Zhong:** Conceptualization, Writing- review & editing, Validation. **Gaowen Chen**: Conceptualization, Writing- review & editing, Validation. **Víctor M. Campello:** Writing- review & editing, Validation. **Karim Lekadir:** Funding acquisition, Project administration, Resources, Supervision, Writing- review & editing, Validation. **Yaosheng Lu**: Conceptualization, Investigation, Data curation, Software, Methodology, Validation, Funding acquisition, Project administration, Resources, Supervision, Writing original draft, Writing– review & editing.

Participants

All the participants: Methodology, Software, Writing– review & editing. **Aloha** (Marawan Elbatel, Robert Martí, Xiaomeng Li), **Angle_avengers** (Gregor Koehler, Raphael Stock, Klaus Maier-Hein), **NKCGP** (Gongping Chen, Lei Zhao, Jianxun Zhang, Yu Dai), **'Kunkunkk'** (Hongkun Sun, Jing Xu), **QiuYaoyang** (Yaoyang Qiu, Panjie Gou), **UCD_Med** (Fangyijie Wang, Guénolé Silvestre, Kathleen Curran), **CQUT-Smart** (Pengzhou Cai, Lu Jiang, Libin Lan) and **aspirerabbit** (Zhanhong Ou).

**Declaration of competing interest**

The authors declare that they have no known competing financial interests or personal relationships that could have appeared to influence the work reported in this paper.

**Acknowledgments**

The organizers would like to thank Yaosheng Lu for the data collection at The First Affiliated Hospital of Jinan University, Mei Zhong for the data collection at the Nanfang Hospital of Southern Medical University and Gaowen Chen for the data collection at the Zhujiang Hospital of Southern Medical University. We would thank these doctors and students for their contributions in data annotation. For target segmentation, annotations were supervised by Gaowen Chen and Di Qiu, under very close advice. We would thank trained raters (i.e., Minghong Zhou, Chao Yuan, Mengqiang Zhou, Xiaosong Jiang, Dengjiang Zhi, Ruiyu Qiu, Zhanhang Song, Sheng Yu, Hao Yi, Hao Liu, Jingbo Rong, Xiaoyan Xie, Jianguo Qi, Zhanhong Ou, Zihao Zhou, Shuangping Chen, Jianmei Jiang, and Yalin Luo) for their contributions in data annotation.

**Funding**

PSFHS Challenge was funded by Natural Science Foundation of Guangdong Province under Grant 2023A1515012833 and 2024A1515011886, Guangzhou Municipal Science and Technology Bureau Guangzhou Key Research and Development Program under Grant 2024B03J1283 and 2024B03J1289, Guangzhou Science and Technology Planning Project under Grant 2023B03J1297, China Scholarship Council under Grant 202206785002, National Key Research and Development Project (2019YFC0120100, 2019YFC0121907, and 2019YFC0121904), and National Natural Science Foundation of China under Grant 61901192.

**Declaration of generative AI**

While preparing this work, the authors used Writefull, DeepL Write, and ChatGPT to enhance the writing structure and re fine grammar. After using these tools, the authors reviewed and edited the content as needed and took full responsibility for the publication's content.

**Appendix: Participation rules and prize policies**

To ensure fairness and transparency in PSFHS Challenge, organizers, data providers, and contributors were prohibited from participating in the challenge since data providers and organizers had access to the data, including the test set ground truth. However, members affiliated with the organizers' institutes were allowed to participate but not eligible for awards and not listed in leaderboard.

Teams receiving a prize had to provide their source codes and present their methodology at MICCAI 2023, sign all necessary prize acceptance documents, and submit a detailed paper outlining their methods. Additionally, participants committed to citing both the data challenge paper and this challenge overview paper in subsequent publications, whether scientific or non-scientific. The challenge results and rankings were publicly announced after the test phase concluded. The top seven teams were awarded a total of 10,000 ¥, with the following distribution: 3000 ¥, 2000 ¥, 1000 ¥, 500 ¥, 500 ¥, 500 ¥ and 500 ¥.

# References

Alsharid, M., Cai, Y., Sharma, H., Drukker, L., Papageorghiou, A.T., Noble, J.A., 2022. Gaze-assisted automatic captioning of fetal ultrasound videos using three-way multi-modal deep neural networks. Medical image analysis 82, 102630.

Angeli, L., Conversano, F., Dall'Asta, A., Volpe, N., Simone, M., Di Pasquo, E., Pignatelli, D., Schera, G.B.L., Di Paola, M., Ricciardi, P., Ferretti, A., Frusca, T., Casciaro, S., Ghi, T., 2020. New technique for automatic sonographic measurement of change in head-perineum distance and angle of progression during active phase of second stage of labor. Ultrasound in obstetrics & gynecology : the official journal of the International Society of Ultrasound in Obstetrics and Gynecology 56, 597-602.

Badr, D.A., Carlin, A., Kadji, C., Kang, X., Cannie, M.M., Jani, J.C., 2024. Timing of induction of labor in suspected macrosomia: retrospective cohort study, systematic review and meta-analysis. Ultrasound in obstetrics & gynecology : the official journal of the International Society of Ultrasound in Obstetrics and Gynecology.

Bai, J., Lu, Y., Liu, H., He, F., Guo, X., 2024. Editorial: New technologies improve maternal and newborn safety. Frontiers in medical technology 6, 1372358.

Bai, J., Sun, Z., Yu, S., Lu, Y., Long, S., Wang, H., Qiu, R., Ou, Z., Zhou, M., Zhi, D., Zhou, M., Jiang, X., Chen, G., 2022. A framework for computing angle of progression from transperineal ultrasound images for evaluating fetal head descent using a novel double branch network. Frontiers in physiology 13, 940150.

Baumgartner, C.F., Kamnitsas, K., Matthew, J., Fletcher, T.P., Smith, S., Koch, L.M., Kainz, B., Rueckert, D., 2017. SonoNet: Real-Time Detection and Localisation of Fetal Standard Scan Planes in Freehand Ultrasound. IEEE transactions on medical imaging 36, 2204-2215.

Cai, Y., Droste, R., Sharma, H., Chatelain, P., Drukker, L., Papageorghiou, A.T., Noble, J.A., 2020. Spatio-temporal visual attention modelling of standard biometry plane-finding navigation. Medical image analysis 65, 101762.

Carneiro, G., Georgescu, B., Good, S., Comaniciu, D., 2008. Detection and measurement of fetal anatomies from ultrasound images using a constrained probabilistic boosting tree. IEEE transactions on medical imaging 27, 1342-1355.

Carvalho Neto, R.H., Viana Junior, A.B., Moron, A.F., Araujo Júnior, E., Carvalho, F.H.C., Feitosa, H.N., 2021. Assessment of the angle of progression and distance perineum-head in the prediction of type of delivery and duration of labor using intrapartum ultrasonography. The journal of maternal-fetal & neonatal medicine : the official journal of the European Association of Perinatal Medicine, the Federation of Asia and Oceania Perinatal Societies, the International Society of Perinatal Obstet 34, 2340-2348.

Chen, G., Bai, J., Ou, Z., Lu, Y., Wang, H., 2024a. PSFHS: Intrapartum ultrasound image dataset for AI-based segmentation of pubic symphysis and fetal head. Scientific data 11, 436.

Chen, G., Li, L., Dai, Y., Zhang, J., Yap, M.H., 2023. AAU-Net: An Adaptive Attention U-Net for Breast Lesions Segmentation in Ultrasound Images. IEEE transactions on medical imaging 42, 1289-1300.

Chen, H., Wu, L., Dou, Q., Qin, J., Li, S., Cheng, J.Z., Ni, D., Heng, P.A., 2017. Ultrasound Standard Plane Detection Using a Composite Neural Network Framework. IEEE transactions on cybernetics 47, 1576-1586.

Chen, Z., Lu, Y., Long, S., Bai, J., 2024b. Dual-path multi-branch feature residual network for salient object detection. Engineering Applications of Artificial Intelligence 133, 108530.

Chen, Z., Lu, Y., Long, S., Campello, V.M., Bai, J., Lekadir, K., 2024c. Fetal Head and Pubic Symphysis Segmentation in Intrapartum Ultrasound Image Using a Dual-Path Boundary-Guided Residual Network. IEEE journal of biomedical and health informatics 28, 4648-4659.

Chen, Z., Ou, Z., Lu, Y., Bai, J., 2024d. Direction-guided and multi-scale feature screening for fetal head–pubic symphysis segmentation and angle of progression calculation. Expert Systems with Applications 245, 123096.

Conversano, F., Peccarisi, M., Pisani, P., Di Paola, M., De Marco, T., Franchini, R., Greco, A., D'Ambrogio, G., Casciaro, S., 2017. Automatic ultrasound technique to measure angle of progression during labor. Ultrasound in obstetrics & gynecology : the official journal of the International Society of Ultrasound in Obstetrics and Gynecology 50, 766-775.

Dall'Asta, A., Angeli, L., Masturzo, B., Volpe, N., Schera, G.B.L., Di Pasquo, E., Girlando, F., Attini, R., Menato, G., Frusca, T., Ghi, T., 2019. Prediction of spontaneous vaginal delivery in nulliparous women with a prolonged second stage of labor: the value of intrapartum ultrasound. American journal of obstetrics and gynecology 221, 642.e641-642.e613.

Fiorentino, M.C., Villani, F.P., Di Cosmo, M., Frontoni, E., Moccia, S., 2023. A review on deep-learning algorithms for fetal ultrasound-image analysis. Medical image analysis 83, 102629.

Ghi, T., Conversano, F., Ramirez Zegarra, R., Pisani, P., Dall'Asta, A., Lanzone, A., Lau, W., Vimercati, A., Iliescu, D.G., Mappa, I., Rizzo, G., Casciaro, S., 2022. Novel artificial intelligence approach for automatic differentiation of fetal occiput anterior and non-occiput anterior positions during labor. Ultrasound in obstetrics & gynecology : the official journal of the International Society of Ultrasound in Obstetrics and Gynecology 59, 93-99.

Ghi, T., Eggebø, T., Lees, C., Kalache, K., Rozenberg, P., Youssef, A., Salomon, L.J., Tutschek, B., 2018. ISUOG Practice Guidelines: intrapartum ultrasound. Ultrasound in obstetrics & gynecology : the official journal of the International Society of Ultrasound in Obstetrics and Gynecology 52, 128-139.

Gimovsky, A.C., 2021. Intrapartum ultrasound for the diagnosis of cephalic malpositions and malpresentations. American journal of obstetrics & gynecology MFM 3, 100438.

Haberman, S., Atallah, F., Nizard, J., Buhule, O., Albert, P., Gonen, R., Ville, Y., Paltieli, Y., 2021. A Novel Partogram for Stages 1 and 2 of Labor Based on Fetal Head Station Measured by Ultrasound: A Prospective Multicenter Cohort Study. American journal of perinatology 38, e14-e20.

Hadad, S., Oberman, M., Ben-Arie, A., Sacagiu, M., Vaisbuch, E., Levy, R., 2021. Intrapartum ultrasound at the initiation of the active second stage of labor predicts spontaneous vaginal delivery. American journal of obstetrics & gynecology MFM 3, 100249.

Huang, Y., Yang, X., Liu, L., Zhou, H., Chang, A., Zhou, X., Chen, R., Yu, J., Chen, J., Chen, C., Liu, S., Chi, H., Hu, X., Yue, K., Li, L., Grau, V., Fan, D.P., Dong, F., Ni, D., 2024. Segment anything model for medical images? Medical image analysis 92, 103061.

Isensee, F., Jaeger, P.F., Kohl, S.A.A., Petersen, J., Maier-Hein, K.H., 2021. nnU-Net: a self-configuring method for deep learning-based biomedical image segmentation. Nature methods 18, 203-211.

Jang, J., Park, Y., Kim, B., Lee, S.M., Kwon, J.Y., Seo, J.K., 2018. Automatic Estimation of Fetal Abdominal Circumference From Ultrasound Images. IEEE journal of biomedical and health informatics 22, 1512-1520.

Kalache, K.D., Dückelmann, A.M., Michaelis, S.A., Lange, J., Cichon, G., Dudenhausen, J.W., 2009. Transperineal ultrasound imaging in prolonged second stage of labor with occipitoanterior presenting fetuses: how well does the 'angle of progression' predict the mode of delivery? Ultrasound in obstetrics & gynecology : the official journal of the International Society of Ultrasound in Obstetrics and Gynecology 33, 326-330.

Kirillov, A., Mintun, E., Ravi, N., Mao, H., Rolland, C., Gustafson, L., Xiao, T., Whitehead, S., Berg, A.C., Lo, W.-Y., 2023. Segment anything, Proceedings of the IEEE/CVF International Conference on Computer Vision, pp. 4015-4026.

LeCun, Y., Bengio, Y., Hinton, G., 2015. Deep learning. Nature 521, 436-444.

Li, L., Hu, Z., Huang, Y., Zhu, W., Wang, Y., Chen, M., Yu, J., 2021. Automatic multi-plaque tracking and segmentation in ultrasonic videos. Medical image analysis 74, 102201.

Li, Y., Liu, Y., Huang, L., Wang, Z., Luo, J., 2022. Deep weakly-supervised breast tumor segmentation in ultrasound images with explicit anatomical constraints. Medical image analysis 76, 102315.

Lin, A., Chen, B., Xu, J., Zhang, Z., Lu, G., Zhang, D., 2022. DS-TransUNet: Dual Swin Transformer U-Net for Medical Image Segmentation. IEEE Transactions on Instrumentation and Measurement 71, 1-15.

Lin, Z., Li, S., Ni, D., Liao, Y., Wen, H., Du, J., Chen, S., Wang, T., Lei, B., 2019. Multi-task learning for quality assessment of fetal head ultrasound images. Medical image analysis 58, 101548.

Lu, Y., Zhi, D., Zhou, M., Lai, F., Chen, G., Ou, Z., Zeng, R., Long, S., Qiu, R., Zhou, M., Jiang, X., Wang, H., Bai, J., 2022a. Multitask Deep Neural Network for the Fully Automatic Measurement of the Angle of Progression. Computational and mathematical methods in medicine 2022, 5192338.

Lu, Y., Zhou, M., Zhi, D., Zhou, M., Jiang, X., Qiu, R., Ou, Z., Wang, H., Qiu, D., Zhong, M., Lu, X., Chen, G., Bai, J., 2022b. The JNU-IFM dataset for segmenting pubic symphysis-fetal head. Data in brief 41, 107904.

Ma, J., He, Y., Li, F., Han, L., You, C., Wang, B., 2024. Segment anything in medical images. Nature communications 15, 654.

Maier-Hein, L., Reinke, A., Godau, P., Tizabi, M.D., Buettner, F., Christodoulou, E., Glocker, B., Isensee, F., Kleesiek, J., Kozubek, M., Reyes, M., Riegler, M.A., Wiesenfarth, M., Kavur, A.E., Sudre, C.H., Baumgartner, M., Eisenmann, M., Heckmann-Nötzel, D., Rädsch, T., Acion, L., Antonelli, M., Arbel, T., Bakas, S., Benis, A., Blaschko, M.B., Cardoso, M.J., Cheplygina, V., Cimini, B.A., Collins, G.S., Farahani, K., Ferrer, L., Galdran, A., van Ginneken, B., Haase, R., Hashimoto, D.A., Hoffman, M.M., Huisman, M., Jannin, P., Kahn, C.E., Kainmueller, D., Kainz, B., Karargyris, A., Karthikesalingam, A., Kofler, F., Kopp-Schneider, A., Kreshuk, A., Kurc, T., Landman, B.A., Litjens, G., Madani, A., Maier-Hein, K., Martel, A.L., Mattson, P., Meijering, E., Menze, B., Moons, K.G.M., Müller, H., Nichyporuk,

B., Nickel, F., Petersen, J., Rajpoot, N., Rieke, N., Saez-Rodriguez, J., Sánchez, C.I., Shetty, S., van Smeden, M., Summers, R.M., Taha, A.A., Tiulpin, A., Tsaftaris, S.A., Van Calster, B., Varoquaux, G., Jäger, P.F., 2024. Metrics reloaded: recommendations for image analysis validation. Nature methods 21, 195-212.

Maier-Hein, L., Reinke, A., Kozubek, M., Martel, A.L., Arbel, T., Eisenmann, M., Hanbury, A., Jannin, P., Müller, H., Onogur, S., Saez-Rodriguez, J., van Ginneken, B., Kopp-Schneider, A., Landman, B.A., 2020. BIAS: Transparent reporting of biomedical image analysis challenges. Medical image analysis 66, 101796.

Mazurowski, M.A., Dong, H., Gu, H., Yang, J., Konz, N., Zhang, Y., 2023. Segment anything model for medical image analysis: An experimental study. Medical image analysis 89, 102918.

Mischi, M., Bell, M.A.L., Van Sloun, R.J., Eldar, Y.C., 2020. Deep learning in medical ultrasound—from image formation to image analysis. IEEE Transactions on Ultrasonics, Ferroelectrics, Frequency Control 67, 2477-2480.

Ou, Z., Bai, J., Chen, Z., Lu, Y., Wang, H., Long, S., Chen, G., 2024. RTSeg-Net: A Lightweight Network for Real-time Segmentation of Fetal Head and Pubic Symphysis from Intrapartum Ultrasound Images. Computers in biology and medicine, 108501.

Pavličev, M., Romero, R., Mitteroecker, P., 2020. Evolution of the human pelvis and obstructed labor: new explanations of an old obstetrical dilemma. American journal of obstetrics and gynecology 222, 3-16.

Pietsch, M., Ho, A., Bardanzellu, A., Zeidan, A.M.A., Chappell, L.C., Hajnal, J.V., Rutherford, M., Hutter, J., 2021. APPLAUSE: Automatic Prediction of PLAcental health via U-net Segmentation and statistical Evaluation. Medical image analysis 72, 102145.

Pu, B., Li, K., Li, S., Zhu, N., 2021. Automatic Fetal Ultrasound Standard Plane Recognition Based on Deep Learning and IIoT. IEEE Transactions on Industrial Informatics 17, 7771-7780.

Qiu, R., Zhou, M., Bai, J., Lu, Y., Wang, H., 2024. PSFHSP-Net: an efficient lightweight network for identifying pubic symphysis-fetal head standard plane from intrapartum ultrasound images. Medical & biological engineering & computing 62, 2975-2986.

Ramirez Zegarra, R., Conversano, F., Dall'Asta, A., Giovanna Di Trani, M., Fieni, S., Morello, R., Melito, C., Pisani, P., Iurlaro, E., Tondo, M., Gabriel Iliescu, D., Nagy, R., Vaso, E., Abou-Dakn, M., Muslu, G., Lau, W., Hung, C., Sirico, A., Lanzone, A., Rizzo, G., Mappa, I., Lees, C., Usman, S., Winkler, A., Braun, C., Levy, R., Vaisbuch, E., Hassan, W.A., Taylor, S., Vimercati, A., Mazzeo, A., Moe Eggebø, T., Amo Wiafe, Y., Ghi, T., Casciaro, S., 2024. A deep learning approach to identify the fetal head position using transperineal ultrasound during labor. European journal of obstetrics, gynecology, and reproductive biology 301, 147-153.

Rueda, S., Fathima, S., Knight, C.L., Yaqub, M., Papageorghiou, A.T., Rahmatullah, B., Foi, A., Maggioni, M., Pepe, A., Tohka, J., Stebbing, R.V., McManigle, J.E., Ciurte, A., Bresson, X., Cuadra, M.B., Sun, C., Ponomarev, G.V., Gelfand, M.S., Kazanov, M.D., Wang, C.W., Chen, H.C., Peng, C.W., Hung, C.M., Noble, J.A., 2014. Evaluation and comparison of current fetal ultrasound image segmentation methods for biometric measurements: a grand challenge. IEEE transactions on medical imaging 33, 797-813.

Ruiping, Y., Kun, L., Shaohua, X., Jian, Y., Zhen, Z., 2024. ViT-UperNet: a hybrid vision transformer with unified-perceptual-parsing network for medical image segmentation. Complex Intelligent Systems, 1-13.

Sharf, Y., Farine, D., Batzalel, M., Megel, Y., Shenhav, M., Jaffa, A., Barnea, O., 2007. Continuous

monitoring of cervical dilatation and fetal head station during labor. Medical engineering & physics 29, 61-71.

Sharma, H., Drukker, L., Chatelain, P., Droste, R., Papageorghiou, A.T., Noble, J.A., 2021. Knowledge representation and learning of operator clinical workflow from full-length routine fetal ultrasound scan videos. Medical image analysis 69, 101973.

Sherer, D.M., 2007. Intrapartum ultrasound. Ultrasound in obstetrics & gynecology : the official journal of the International Society of Ultrasound in Obstetrics and Gynecology 30, 123-139.

Torres, H.R., Morais, P., Oliveira, B., Birdir, C., Rüdiger, M., Fonseca, J.C., Vilaça, J.L., 2022. A review of image processing methods for fetal head and brain analysis in ultrasound images. Computer methods and programs in biomedicine 215, 106629.

Tutschek, B., Torkildsen, E.A., Eggebø, T.M., 2013. Comparison between ultrasound parameters and clinical examination to assess fetal head station in labor. Ultrasound in obstetrics & gynecology : the official journal of the International Society of Ultrasound in Obstetrics and Gynecology 41, 425-429.

van den Heuvel, T.L.A., de Bruijn, D., de Korte, C.L., Ginneken, B.V., 2018. Automated measurement of fetal head circumference using 2D ultrasound images. PloS one 13, e0200412.

Vesal, S., Gayo, I., Bhattacharya, I., Natarajan, S., Marks, L.S., Barratt, D.C., Fan, R.E., Hu, Y., Sonn, G.A., Rusu, M., 2022. Domain generalization for prostate segmentation in transrectal ultrasound images: A multi-center study. Medical image analysis 82, 102620.

Vogel, J.P., Pujar, Y., Vernekar, S.S., Armari, E., Pingray, V., Althabe, F., Gibbons, L., Berrueta, M., Somannavar, M., Ciganda, A., Rodriguez, R., Bendigeri, S., Kumar, J.A., Patil, S.B., Karinagannanavar, A., Anteen, R.R., Mallappa Ramachandrappa, P., Shetty, S., Bommanal, L., Haralahalli Mallesh, M., Gaddi, S.S., Chikkagowdra, S., Raghavendra, B., Homer, C.S.E., Lavender, T., Kushtagi, P., Hofmeyr, G.J., Derman, R., Goudar, S., 2024. Effects of the WHO Labour Care Guide on cesarean section in India: a pragmatic, stepped-wedge, cluster-randomized pilot trial. Nature medicine 30, 463-469.

Wang, H., Xie, S., Lin, L., Iwamoto, Y., Han, X.-H., Chen, Y.-W., Tong, R., 2022a. Mixed transformer u-net for medical image segmentation, ICASSP 2022-2022 IEEE international conference on acoustics, speech and signal processing (ICASSP). IEEE, pp. 2390-2394.

Wang, X., Wang, W., Cai, X., 2022b. Automatic measurement of fetal head circumference using a novel GCN-assisted deep convolutional network. Computers in biology and medicine 145, 105515.

Wiesenfarth, M., Reinke, A., Landman, B.A., Eisenmann, M., Saiz, L.A., Cardoso, M.J., Maier-Hein, L., Kopp-Schneider, A., 2021. Methods and open-source toolkit for analyzing and visualizing challenge results. Scientific reports 11, 2369.

Wright, R., Gomez, A., Zimmer, V.A., Toussaint, N., Khanal, B., Matthew, J., Skelton, E., Kainz, B., Rueckert, D., Hajnal, J.V., Schnabel, J.A., 2023. Fast fetal head compounding from multi-view 3D ultrasound. Medical image analysis 89, 102793.

Wu, L., Cheng, J.Z., Li, S., Lei, B., Wang, T., Ni, D., 2017. FUIQA: Fetal Ultrasound Image Quality Assessment With Deep Convolutional Networks. IEEE transactions on cybernetics 47, 1336-1349.

Yang, X., Yu, L., Li, S., Wen, H., Luo, D., Bian, C., Qin, J., Ni, D., Heng, P.A., 2019. Towards Automated Semantic Segmentation in Prenatal Volumetric Ultrasound. IEEE transactions on medical imaging 38, 180-193.

Yu, Z., Tan, E.L., Ni, D., Qin, J., Chen, S., Li, S., Lei, B., Wang, T., 2018. A Deep Convolutional Neural Network-Based Framework for Automatic Fetal Facial Standard Plane Recognition. IEEE journal of biomedical and health informatics 22, 874-885.

Zamzmi, G., Rajaraman, S., Hsu, L.Y., Sachdev, V., Antani, S., 2022. Real-time echocardiography image analysis and quantification of cardiac indices. Medical image analysis 80, 102438.

Zhao, H., Zheng, Q., Teng, C., Yasrab, R., Drukker, L., Papageorghiou, A.T., Noble, J.A., 2023. Memory-based unsupervised video clinical quality assessment with multi-modality data in fetal ultrasound. Medical image analysis 90, 102977.

Zhou, M., Wang, C., Lu, Y., Qiu, R., Zeng, R., Zhi, D., Jiang, X., Ou, Z., Wang, H., Chen, G., Bai, J., 2023. The segmentation effect of style transfer on fetal head ultrasound image: a study of multi-source data. Medical & biological engineering & computing 61, 1017-1031.

Zhu, L., Wang, X., Ke, Z., Zhang, W., Lau, R.W., 2023. Biformer: Vision transformer with bi-level routing attention, Proceedings of the IEEE/CVF conference on computer vision and pattern recognition, pp. 10323-10333.

Zimmer, V.A., Gomez, A., Skelton, E., Wright, R., Wheeler, G., Deng, S., Ghavami, N., Lloyd, K., Matthew, J., Kainz, B., Rueckert, D., Hajnal, J.V., Schnabel, J.A., 2023. Placenta segmentation in ultrasound imaging: Addressing sources of uncertainty and limited field-of-view. Medical image analysis 83, 102639.